\documentclass[nofootinbib,aps,12pt]{revtex4-1}
\usepackage{graphicx}
\usepackage{hyperref}
\usepackage{cancel}
\usepackage{amssymb}
\usepackage{textcomp}
\usepackage{amsmath}
\usepackage{amsfonts}
\usepackage{bm}
\usepackage{times}
\usepackage{epsfig}
\usepackage{color}
\usepackage{mathptmx}
\usepackage[normalem]{ulem}
\usepackage{ mathrsfs }
\usepackage{textcomp}
\usepackage{slashed}
\usepackage{booktabs}
\usepackage{latexsym}
\usepackage{verbatim}
\usepackage{extarrows}
\usepackage{multirow}
\usepackage{xcolor}
\hypersetup{
	colorlinks,
	linkcolor={red!50!black},
	citecolor={blue!50!black},
	urlcolor={blue!80!black},
	bookmarksopen=true
}
\newcommand{\GeV}{{\rm GeV}}
\newcommand{\TeV}{{\rm TeV}}
\newcommand{\fb}{{\rm fb}}
\newcommand{\tr}{{\rm tr}}

\makeatletter
\renewcommand*\env@matrix[1][\arraystretch]{%
	\edef\arraystretch{#1}%
	\hskip -\arraycolsep
	\let\@ifnextchar\new@ifnextchar
	\array{*\c@MaxMatrixCols c}}
\makeatother
\begin{document}
\title{\LARGE Higgs production at future $e^+e^-$ colliders in the Georgi-Machacek model}
\bigskip

\author{Bin Li$^{1}$}
\email{Current address: School of Physics, Peking University, Beijing 100871, China; libin@pku.edu.cn}
\author{Zhi-Long Han$^{2}$}
\email{sps\_hanzl@ujn.edu.cn}
\author{Yi Liao$^{1,3,4}$}
\email{liaoy@nankai.edu.cn}

\affiliation{
	$^1$~School of Physics, Nankai University, Tianjin 300071, China
	\\
    $^2$ School of Physics and Technology, University of Jinan, Jinan, Shandong 250022, China
	\\
	$^3$ CAS Key Laboratory of Theoretical Physics, Institute of Theoretical Physics,
	Chinese Academy of Sciences, Beijing 100190, China
	\\
	$^4$ Center for High Energy Physics, Peking University, Beijing 100871, China
}
\date{\today}

\begin{abstract}
We study how the dominant single and double SM-like Higgs ($h$) production at future $e^+e^-$ colliders is modified in the Georgi-Machacek (GM) model. On imposing theoretical, indirect and direct constraints, significant deviations of $h$-couplings from their SM values are still possible; for instance, the Higgs-gauge coupling coupling can be corrected by a factor $\kappa_{hVV}\in[0.93,1.15]$ in the allowed parameter space. For the Higgs-strahlung $e^+e^-\to hZ$ and vector boson fusion processes $e^+e^-\to h\nu\bar{\nu},~he^+e^-$, the cross section could increase by $32\%$ or decrease by $13\%$. In the case of associated production with a top quark pair $e^+e^-\to ht\bar{t}$, the cross section can be enhanced up to several times when the custodial triplet scalar $H_3^0$ is resonantly produced. In the meanwhile, the double Higgs production $e^+e^-\to hhZ~(hh\nu\bar{\nu})$ can be maximally enhanced by one order of magnitude at the resonant $H_{1,3}^0$ production.  We also include exclusion limits expected from future LHC runs at higher energy and luminosity and discuss their further constraints on the relevant model parameters. We find that  the GM model can result in likely measurable deviations of Higgs production from the SM at future $e^+e^-$ colliders.
\end{abstract}

\graphicspath{{figure/}}
\maketitle
\baselineskip=16pt

\pagenumbering{arabic}

\vspace{1.0cm}
\tableofcontents

\newpage

\section{Introduction}
\label{sec:introduction}

The discovery of a 125~GeV scalar at the Large Hadron Collider (LHC)~\cite{Aad:2012tfa,Chatrchyan:2012xdj} confirmed the Higgs mechanism of electroweak symmetry breaking in the standard model (SM)~\cite{Englert:1964et,Higgs:1964ia,Higgs:1964pj,Guralnik:1964eu,Higgs:1966ev,Kibble:1967sv}. Yet the sector that triggers the symmetry breaking remains to be disclosed.
With an elementary scalar the SM is also confronted with issues such as gauge hierarchy and flavor problems. To solve or relax some of these issues, there have been various theoretical attempts beyond the SM whose phenomenologies have been extensively studied, such as in the minimal supersymmetric model (MSSM)~\cite{Nilles:1983ge,Haber:1984rc,Djouadi:2005gj,
Cao:2012fz}, two Higgs doublet models (THDM)~\cite{Ma:2000cc,Davidson:2009ha,Haba:2011nb,Branco:2011iw,
Wang:2016vfj,Guo:2017ybk}, little Higgs models~\cite{ArkaniHamed:2001nc,ArkaniHamed:2002pa,
ArkaniHamed:2002qx,Low:2002ws,Han:2003wu}, and composite Higgs models~\cite{Agashe:2004rs,Gripaios:2009pe,Marzocca:2012zn}, to mention a few with a modified Higgs sector.

To unravel the symmetry breaking sector it is necessary to measure the interactions of the Higgs boson with itself and other related particles. While this is believed to be very challenging at the LHC~\cite{Duhrssen:2004cv}, future lepton colliders provide an avenue to study these interactions at a reasonably good precision due to cleaner environment~\cite{Lafaye:2017kgf}. There have been several proposals for next-generation $e^+e^-$ colliders that are under active studies, including the Circular Electron-Positron Collider (CEPC)~\cite{CEPC-SPPCStudyGroup:2015csa,CEPC-SPPCStudyGroup:2015esa}, the Future Circular Collider (FCC-ee)~\cite{Gomez-Ceballos:2013zzn}, the Compact Linear Collider (CLIC)~\cite{Aicheler:2012bya,Abramowicz:2016zbo}, and the International Linear Collider (ILC)~\cite{Baer:2013cma,Asner:2013psa,Fujii:2015jha}. These colliders are planned to operate at a center of mass (CM) energy ranging from about 250~GeV to 3~TeV, thereby making accessible most of dominant production processes of the Higgs boson.

At an $e^+e^-$ collider of high enough energy one could study simultaneously the interactions of the Higgs boson $h$ with itself and with gauge bosons $W^\pm,~Z$ or even fermions such as the top quark $t$. This would be very helpful for us to build an overall picture on the symmetry breaking sector and gain a hint on possible physics that goes beyond the SM. The $hZZ$ coupling can be measured via the Higgs-strahlung process $e^+e^-\to hZ$ and the $ZZ$ fusion process $e^+e^-\to (ZZ\to h)e^+e^-$, while the $hWW$ coupling can be probed via the $WW$ fusion process $e^+e^-\to (WW\to h)\nu\bar{\nu}$. The top Yukawa coupling can possibly be extracted from the associated production process $e^+e^-\to ht\bar{t}$. And finally, the Higgs pair production processes $e^+e^-\to hh\nu\bar{\nu},~hhZ$ provide an access to the trilinear coupling of the Higgs boson. These processes are generally modified by new interactions or new heavy particles, and precise measurements on them could help us identify the imprints of physics beyond the SM~\cite{Yue:2003yk,Yue:2005av,Liu:2006rc,Arhrib:2008jp,LopezVal:2009qy,Asakawa:2010xj,Heng:2013wia,
Yang:2014tia,Yang:2014gca,Liu:2014uua,Han:2015orc,Antusch:2015gjw,Kanemura:2016tan,Khosa:2016jly,
DeCurtis:2017gzi,Guo:2017ugk,Gu:2017ckc}.

In this work we will study the above mentioned dominant single and double Higgs production at future $e^+e^-$ colliders in the Georgi-Machacek (GM) model \cite{Georgi:1985nv,Chanowitz:1985ug}. The model is interesting because it introduces weak isospin-triplet scalars in a manner that preserves the custodial $SU(2)_V$ symmetry. While this symmetry guarantees that the $\rho$ parameter is naturally unity at the tree level, the arrangement of the triplet scalars allows them to develop a vacuum expectation value (VEV) as large as a few tens of GeV. After spontaneous symmetry breaking there remain ten physical scalars that can be approximately classified into irreducible representations of $SU(2)_V$, one quintuplet, one triplet and two singlets. These multiplets couple to gauge bosons and fermions differently. In particular, the couplings of the SM-like Higgs boson $h$ can be significantly modified, and processes involving $h$ receive additional contributions from new scalars as intermediate states. It is a distinct feature of the GM model compared to a scalar sector with only singlet and doublet scalars that the $h$ couplings to fermions and gauge bosons may be enhanced, or always enhanced in the case of the quartic couplings to a gauge boson pair.

The GM model has been extended by embedding it in more elaborate theoretical scenarios such as little Higgs~\cite{Chang:2003un,Chang:2003zn} and supersymmetric models~\cite{Cort:2013foa,Vega:2017gkk}, by generalizing it to larger $SU(2)$ multiplets~\cite{Logan:2015xpa} or including dark matter~\cite{Campbell:2016zbp,Pilkington:2017qam}. The phenomenology of exotic scalars has previously been studied, including searches for exotic scalars and the application of a variety of constraints on the model parameter space~\cite{Haber:1999zh,Godfrey:2010qb,Logan:2010en,Chang:2012gn,
Chiang:2012cn,Kanemura:2013mc,Englert:2013zpa,Englert:2013wga,Chiang:2013rua,Efrati:2014uta,
Chiang:2014hia,Godunov:2015lea,Chiang:2015rva,Chiang:2015amq,Degrande:2015xnm,Arroyo-Urena:2016gjt,
Blasi:2017xmc,Zhang:2017och,Chiang:2017vvo,Degrande:2017naf,Logan:2017jpr,Krauss:2017xpj,Sun:2017mue}; in particular, previous works on $e^+e^-$ colliders~\cite{Chiang:2015rva,Zhang:2017och} mainly concentrated on the custodial quintuplet particles. When these exotic scalars are heavy, it is difficult to produce them directly even at LHC, but we will show that they could be probed indirectly at $e^+e^-$ colliders via modifications to the SM-like Higgs production processes. If the new scalars are light enough, they could contribute as resonances in those processes and thus affect them more significantly. In either case, high energy $e^+e^-$ colliders could provide a viable way to test the GM model.

This paper is organized as follows. We recall in the next section the basic features in the Higgs sector of the Georgi-Machacek model. We discuss in Section~\ref{sec:constraints} various constraints on the model parameter space coming from current and future LHC runs as well as from low energy precision measurements and theoretical considerations. Then we investigate various SM-like Higgs production processes in Section~\ref{sec:phenomenology} at the $500~\GeV$ and $1~\TeV$ ILC. We reserve for our future work a comparative study of electron colliders operating at various energies and luminosities which requires a detailed simulation of the relevant processes. We finally summarize our main results in Section~\ref{sec:conclusion}. Some lengthy coefficients and Higgs trilinear couplings are delegated to Appendixes~\ref{app:htt}, \ref{app:trilinear}, and \ref{app:hhvv}.

\section{The Georgi-Machacek Model}
\label{sec:GM model}

The model contains the usual Higgs doublet scalar $\phi$ with hypercharge $Y=1/2$ and introduces a new complex triplet scalar $\chi$ with $Y=1$ and a new real triplet scalar $\xi$ with $Y=0$. To make the custodial symmetry manifest, it is convenient to recast them in a matrix form:
\begin{align}
\Phi=
\begin{pmatrix}
\phi^{0*}&\phi^+\\
-\phi^-   &\phi^0
\end{pmatrix},~
\Delta=
\begin{pmatrix}
\chi^{0*}&\xi^+ &\chi^{++}\\
-\chi^-  &\xi^0 &\chi^+\\
\chi^{--}&-\xi^-&\chi^0
\end{pmatrix},
\end{align}
using the phase convention, $\chi^{--}=(\chi^{++})^*,~\chi^-=(\chi^+)^*,~\xi^-=(\xi^+)^*,~\phi^-=(\phi^+)^*$, and $\xi^0=(\xi^0)^*$. The matrices $\Phi$ and $\Delta$ transform under $SU(2)_L\times SU(2)_R$ as $\Phi\to U_L\Phi U_R^\dagger$ and $\Delta\to U_L\Delta U_R^\dagger$ with $U_{L,R}=\exp(i\theta_{L,R}^a T^a)$, where, for $\Phi$, $T^a=\tau^a/2$ with $\tau^a$ being the Pauli matrices, and for $\Delta$, $T^a=t^a$ are
\begin{align}
t^1=\frac{1}{\sqrt{2}}
\begin{pmatrix}
0&~~1&~~0\\
1&~~0&~~1\\
0&~~1&~~0
\end{pmatrix},~
t^2=\frac{1}{\sqrt{2}}
\begin{pmatrix}
0&~~-i&~~0\\
i&~~0&~~-i\\
0&~~i&~~0
\end{pmatrix},~
t^3=
\begin{pmatrix}
1&~~0&~~0\\
0&~~0&~~0\\
0&~~0&~~-1
\end{pmatrix}.
\end{align}

The most general scalar potential invariant under $SU(2)_L\times SU(2)_R\times U(1)_Y$ is given by~\cite{Hartling:2014zca}
\begin{align}
\nonumber
V_H=&
\frac{1}{2}\mu_2^2\tr(\Phi^\dagger \Phi)+\frac{1}{2}\mu_3^2\tr(\Delta^\dagger \Delta)+\lambda_1[\tr(\Phi^\dagger\Phi)]^2+\lambda_2\tr(\Phi^\dagger\Phi)\tr(\Delta^\dagger \Delta)
\\
\nonumber
&+\lambda_3\tr(\Delta^\dagger \Delta \Delta^\dagger \Delta)+\lambda_4[\tr(\Delta^\dagger \Delta)]^2-\lambda_5\tr(\Phi^\dagger \tau^a \Phi \tau^b)\tr(\Delta^\dagger t^a \Delta t^b)
\\
&-M_1\tr(\Phi^\dagger\tau^a\Phi\tau^b)(U\Delta U^\dagger)_{ab}-M_2\tr(\Delta^\dagger t^a \Delta t^b)(U \Delta U^\dagger)_{ab},
\end{align}
where all free parameters are real and the matrix $U$ is~\cite{Aoki:2007ah}
\begin{equation}
U=
\begin{pmatrix}
-\frac{1}{\sqrt{2}}&~~0&~~\frac{1}{\sqrt{2}}\\
-\frac{i}{\sqrt{2}}&~~0&~~-\frac{i}{\sqrt{2}}\\
0&~~1&~~0
\end{pmatrix}.
\end{equation}
The spontaneous symmetry breaking is triggered by the VEVs $\langle\Phi\rangle=1_2v_\phi/\sqrt{2}$ and $\langle\Delta\rangle=1_3v_\Delta$. As usual, the weak gauge bosons obtain masses from the kinetic terms of the scalars
\begin{equation}
\label{equ:lkin}
\mathcal{L}=\frac{1}{2}\tr[(D^\mu \Phi)^\dagger D_\mu \Phi]+\frac{1}{2}\tr[(D^\mu \Delta)^\dagger D_\mu \Delta],
\end{equation}
where
\begin{align}
\nonumber
D_\mu\Phi &= \partial_\mu \Phi + ig_2 A_\mu^a \frac{\tau^a}{2}\Phi
-ig_1B_\mu\Phi \frac{\tau^3}{2},
\\
D_\mu\Delta &= \partial_\mu \Delta + ig_2 A_\mu^a t^a\Delta
-ig_1B_\mu\Delta t^3,
\end{align}
with $g_{2,1}$ being the gauge couplings of $SU(2)_L\times U(1)_Y$.
Their squared masses are $m_W^2=g_2^2v^2/4$ and $m_Z^2=(g_1^2+g_2^2)v^2/4$ where
\begin{align}
v^2=v_\phi^2+8v_\Delta^2,
\end{align}
which should be identified with $1/(\sqrt{2}G_F)$ where $G_F$ is the Fermi constant. The parameter $\rho=1$ is thus established at the tree level. Since the custodial symmetry is explicitly broken by hypercharge and Yukawa couplings, divergent radiative corrections to $\rho$ will generally appear at one loop within the framework of the GM model~\cite{Gunion:1990dt}.

Restricting our discussions to the tree level, the custodial symmetry is respected by the scalar spectra so that the scalars can be classified into irreducible representations of $SU(2)_V$. Excluding the would-be Nambu-Goldstone bosons to be eaten up by the weak gauge bosons, they are decomposed into a quintuplet, a triplet and two singlets. Denoting the real and imaginary parts of the neutral components of the original fields after extracting out the VEVs,
\begin{align}
\phi^0=\frac{1}{\sqrt{2}}(v_\phi+\phi^r+i\phi^i),~
\chi^0=v_\Delta+\frac{1}{\sqrt{2}}(\chi^r+i\chi^i),~
\xi^0=v_\Delta+\xi^r,
\end{align}
the quintuplet states are~\cite{Hartling:2014zca}
\begin{align}
H_5^{++}&=\chi^{++},
\nonumber
\\
H_5^+&=\frac{1}{\sqrt{2}}(\chi^+-\xi^+),
\nonumber
\\
H_5^0&=\sqrt{\frac{2}{3}}\xi^r-\sqrt{\frac{1}{3}}\chi^r,
\end{align}
plus $H_5^{--,-}=(H_5^{++,+})^*$. The triplet states are
\begin{align}
H_3^+&=-s_H\phi^++c_H\frac{1}{\sqrt{2}}(\chi^++\xi^+),
\nonumber
\\
H_3^0&=-s_H\phi^i+c_H\chi^i,
\end{align}
plus $H_3^-=(H_3^+)^*$, where the doublet-triplet mixing angle $\theta_H$ is defined by
\begin{align}
\label{eqn:chsh}
c_H\equiv\cos \theta_H=\frac{v_\phi}{v},~
s_H\equiv\sin \theta_H=\frac{2\sqrt{2}v_\Delta}{v}.
\end{align}
We denote the quintuplet and triplet masses as $m_5$ and $m_3$ respectively. At the Lagrangian level, the quintuplet scalars couple to the electroweak gauge bosons but not to fermions (i.e., with $H_5VV$ but without $H_5f\bar{f}$ couplings), while the opposite is true for the triplet scalars (with $H_3f\bar{f}$ but without $H_3VV$ couplings).

The two custodial singlets mix by an angle $\alpha$ into the mass eigenstates
\begin{align}
h&=c_\alpha \phi^r-s_\alpha \left(\sqrt{\frac{1}{3}}\xi^r+\sqrt{\frac{2}{3}}\chi^r\right),
\nonumber
\\
H_1^0&=s_\alpha \phi^r + c_\alpha \left(\sqrt{\frac{1}{3}}\xi^r+\sqrt{\frac{2}{3}}\chi^r\right),
\end{align}
with $c_\alpha=\cos\alpha$ and $s_\alpha=\sin\alpha$. We assume that the lighter state $h$ is the observed 125~GeV scalar~\cite{Aad:2012tfa,Chatrchyan:2012xdj,Aad:2015zhl}. The angle is determined by~\cite{Hartling:2014zca}
\begin{align}\label{eq:mixing_angle_singlet}
\sin 2\alpha =\frac{2M_{12}^2}{m_{1}^2-m_h^2},~
\cos 2\alpha =\frac{M_{22}^2-M_{11}^2}{m_{1}^2-m_h^2},
\end{align}
where $m_{h,1}$ are the masses of $h$ and $H_1^0$ respectively, and in terms of the scalar couplings and VEVs,
\begin{align}
\nonumber
M_{11}^2=&8\lambda_1v_\phi^2,
\\
\nonumber
M_{12}^2=&\frac{\sqrt{3}}{2}v_\phi\left[-M_1+4(2\lambda_2-\lambda_5)v_\Delta\right],\\
M_{22}^2=&\frac{M_1v_\phi^2}{4v_\Delta}-6M_2v_\Delta+8(\lambda_3+3\lambda_4)v_\Delta^2.
\end{align}

\section{Constraints on Parameter Space}
\label{sec:constraints}

\begin{figure}
   \centering
	\includegraphics[width=0.45\linewidth]{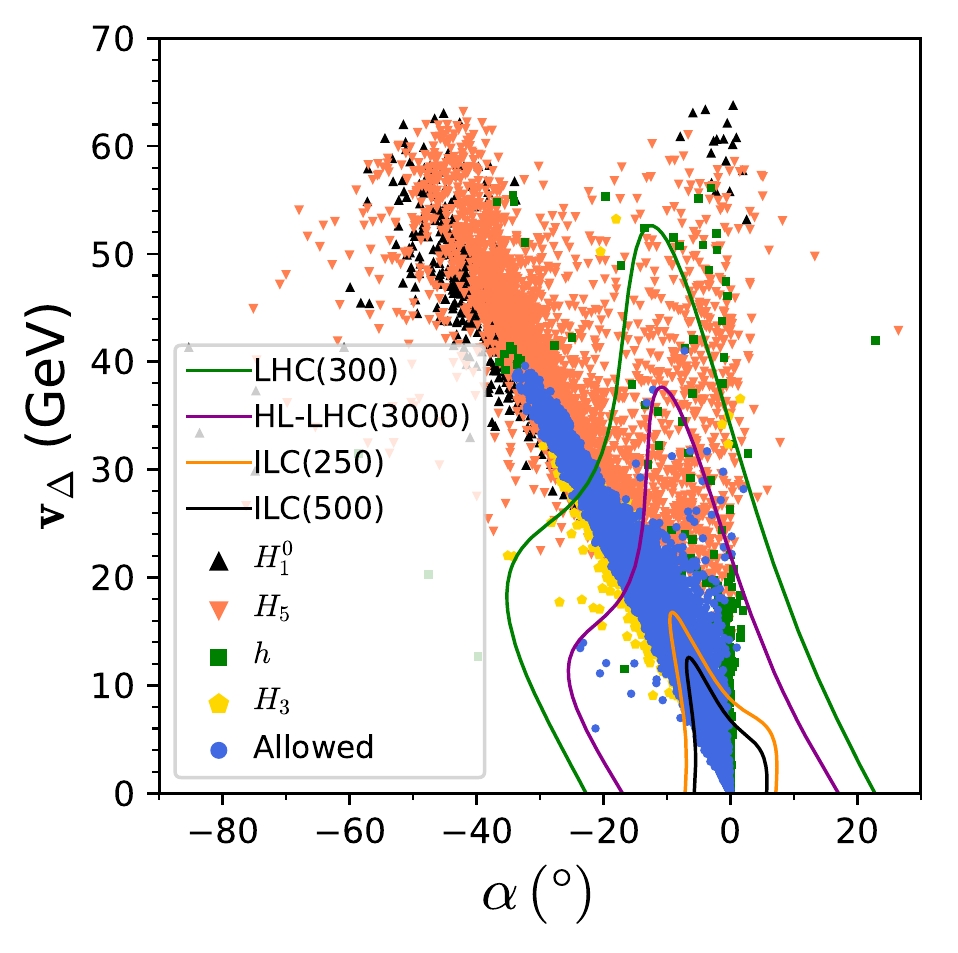}
	\caption{Distribution of survived points in the $\alpha-v_\Delta$ plane under theoretical and indirect constraints generated by GMCALC. The black, red, yellow, and green points are further excluded by direct searches for $H_1^0$, $H_5$, and $H_3$ and by Higgs signal strength analysis respectively, while the circular blue points pass all current constraints. The green, magenta, orange, and black curves are exclusion bounds expected from projection results of 14 TeV LHC with $300~\fb^{-1}$, HL-LHC with $3000~\fb^{-1}$, and ILC at 250 and 500 GeV.}
	\label{fig:constraints}
\end{figure}

There are generally many free parameters in the scalar potential with an extended scalar sector. In this section, we will perform a combined analysis based on theoretical considerations and indirect and direct constraints to obtain the allowed regions for the parameters that are most relevant for our later Higgs production processes. The theoretical constrains are mainly derived from the requirement of perturbativity and vacuum stability~\cite{Aoki:2007ah,Hartling:2014zca}, while the indirect ones cover the oblique parameters ($S,~T,~U$), the $Z$-pole observables ($R_b$), and the $B$-meson observables~\cite{Hartling:2014aga,Hartling:2014zca}. Among the $B$-meson observables ($B_s^0-\bar{B}_s^0$ mixing, $B_s^0\to \mu^+\mu^-$, and $b\to s\gamma$), the decay $b\to s\gamma$ currently sets the strongest bound. All of these constraints have been implemented in the calculator GMCALC~\cite{Hartling:2014xma} for the GM model, which will be applied as our starting point. On top of this we will impose up-to-date direct experimental constraints which cover the searches for a heavy neutral Higgs boson ($H_1^0$), custodial triplet bosons ($H_3$) and quintuplet bosons ($H_5$), and the signal strength analysis of the SM-like Higgs boson ($h$). For the signal strength analysis, we will also include the constraints expected from future runs of LHC and the proposed ILC. In Fig.~\ref{fig:constraints} we show how survived points in the $\alpha-v_\Delta$ plane evolve with the inclusion of various constraints. With theoretical and indirect constraints alone, a $v_\Delta$ as large as $60~\GeV$ is still allowed. When the constraints from direct searches for $H_1^0$ (black points), $H_5$ (red), $H_3$ (yellow) and from the Higgs signal strength (green) are included, more and more points are excluded. At this stage, we have $v_\Delta\lesssim40~\GeV$, and $\alpha<0$ is preferred. Also shown in the figure are the future prospects of constraints derived from Higgs signal strength measurements at 14 TeV LHC with $300~\fb^{-1}$ (green curve), High-Luminosity LHC (HL-LHC) with $3000~\fb^{-1}$ (magenta)~\cite{ATLAS:HL-LHC}, ILC with $1150~\fb^{-1}$ at $250~\GeV$ (orange) and $1600~\fb^{-1}$ at $500~\GeV$ (black)~\cite{Asner:2013psa}. It is clear that a wide parameter space will be within the reach of future ILC operations. In the following subsections we will describe these direct experimental constraints in more detail.

\subsection{Singlet Searches}

\begin{figure}
    \centering
	\includegraphics[width=0.45\linewidth]{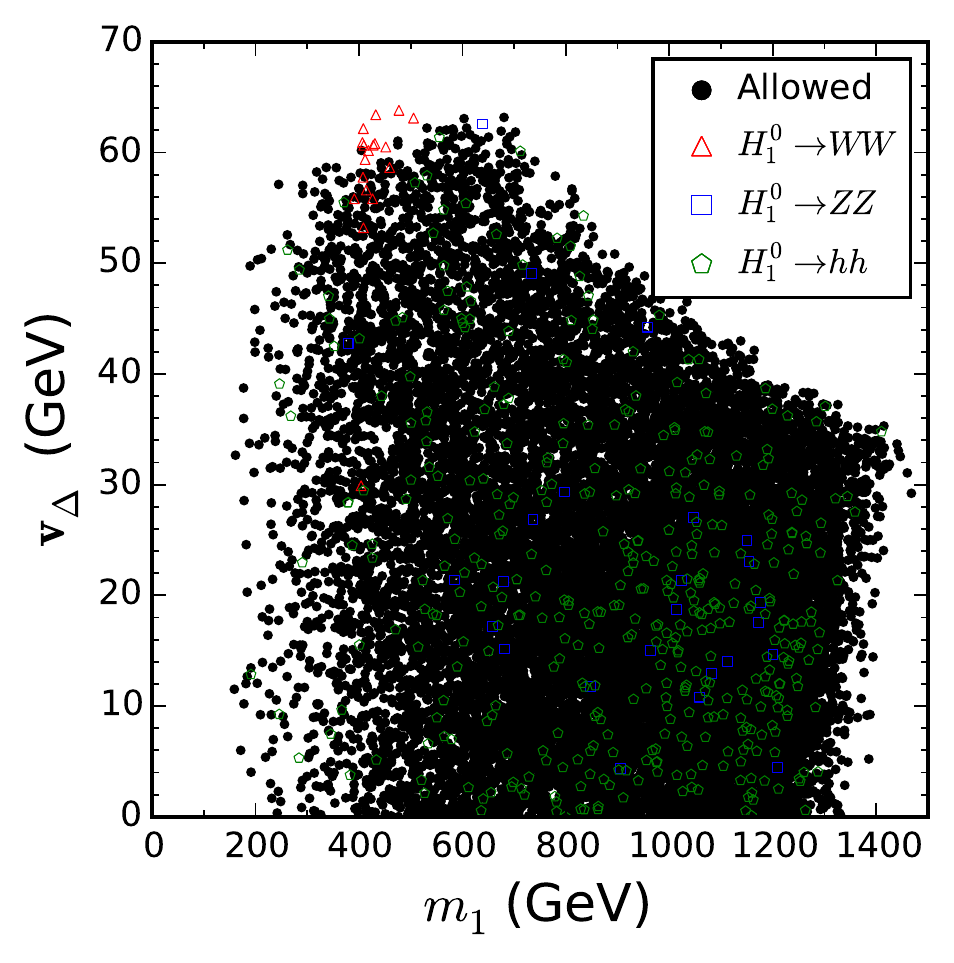}~
    \includegraphics[width=0.465\linewidth]{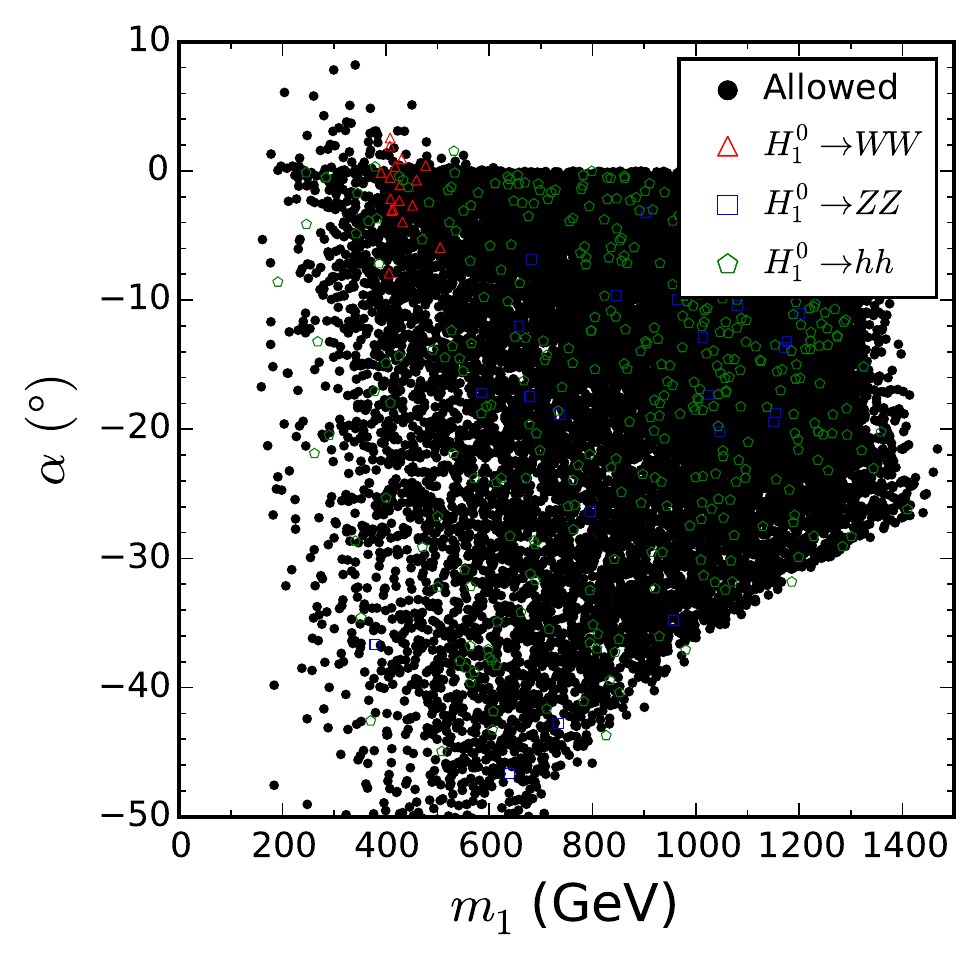}
	\caption{Distribution of survived points shown in the $v_\Delta-m_1$ plane (left) and $\alpha-m_1$ plane (right). The red, blue, and green points are excluded by the $H_1^0\to WW,
~ZZ,~hh$ searches respectively.}
	\label{fig:H1}
\end{figure}

In the GM model, the heavy neutral Higgs boson $H_1^0$ can decay to a pair of vector bosons when it is above the threshold. Searches for heavy resonances decaying to a $WW$~\cite{ATLAS:2016kjy} and $ZZ$~\cite{ATLAS:2016npe} pair are performed by the ATLAS Collaboration using the data collected at $\sqrt{s}=13~\TeV$ with an integrated luminosity of $13.2~\fb^{-1}$. For the gluon fusion (ggF) and vector boson fusion (VBF) production of the heavy Higgs, the corresponding cross sections are calculated as
\begin{align}
\label{equ:ggf}
\nonumber
\sigma_{\text{ggF}}^{\text{GM}}=&\sigma_{\text{The}}(gg\to H_1^0)\times\kappa_{H_1^0f\bar{f}}^2\times \text{BR}_{\text{GM}}(H_1^0\to VV),\\
\sigma_{\text{VBF}}^{\text{GM}}=&\sigma_{\text{The}}(qq\to H_1^0)\times\kappa_{H_1^0VV}^2\times \text{BR}_{\text{GM}}(H_1^0\to VV),
\end{align}
where
\begin{align}
\label{equ:heavy higgs}
\nonumber
\kappa_{H_1^0f\bar{f}}&\equiv \frac{g_{H_1^0f\bar{f}}}{g_{hf\bar{f}}^\text{SM}}
=\frac{s_\alpha}{c_H},
\\
\kappa_{H_1^0VV}&\equiv \frac{g_{H_1^0VV}}{g_{hVV}^\text{SM}}
=s_\alpha c_H+\frac{2\sqrt{6}}{3}c_\alpha s_H,
\end{align}
with $g$s denoting the couplings in the SM and the GM model. The theoretical cross sections of a SM-like heavy Higgs $\sigma_{\text{The}}(gg\to H_1^0)$ and $\sigma_{\text{The}}(qq\to H_1^0)$ have been tabulated in Ref.~\cite{Heinemeyer:2013tqa}, while the branching ratio $\text{BR}_{\text{GM}}(H_1^0\to VV)$ is obtained using GMCALC.

When $m_1>2m_h$, $H_1^0$ can also decay into a Higgs pair $hh$, which may greatly enhance the Higgs pair production at the LHC. The cross section for resonant production of a Higgs boson pair is given by~\cite{Chang:2017niy}
\begin{equation}
\sigma(pp\to H_1^0 \to hh) = \sigma_{\text{The}}(gg\to H_1^0)\times\kappa_{H_1^0f\bar{f}}^2\times \text{BR}_{\text{GM}}(H_1^0\to hh),
\end{equation}
where $\text{BR}_{\text{GM}}(H_1^0\to hh)$ is also calculated by GMCALC. Recently, a search for resonant Higgs boson pair production ($H_1^0\to hh$) has been performed by the CMS Collaboration~\cite{Sirunyan:2017guj} in the $b\bar{b}\ell\ell \nu\nu$ final state.

In Fig.~\ref{fig:H1} we show in the $v_\Delta-m_1$ and $\alpha-m_1$ planes the survived points upon applying the constraints from the direct searches $H_1^0\to WW$, $H_1^0\to ZZ$, and $H_1^0\to hh$. While the $H_1^0\to WW,~ZZ$ searches remove only a small portion of points, a considerable portion is excluded by the $H_1^0\to hh$ search. Due to large variations of BR($H_1^0\to hh$) in our scan~\cite{Chang:2017niy}, no clear dependence of the exclusion bounds on $v_\Delta$ and $\alpha$ is visible.

\subsection{Triplet Searches}

\begin{figure}
    \centering
	\includegraphics[width=0.45\linewidth]{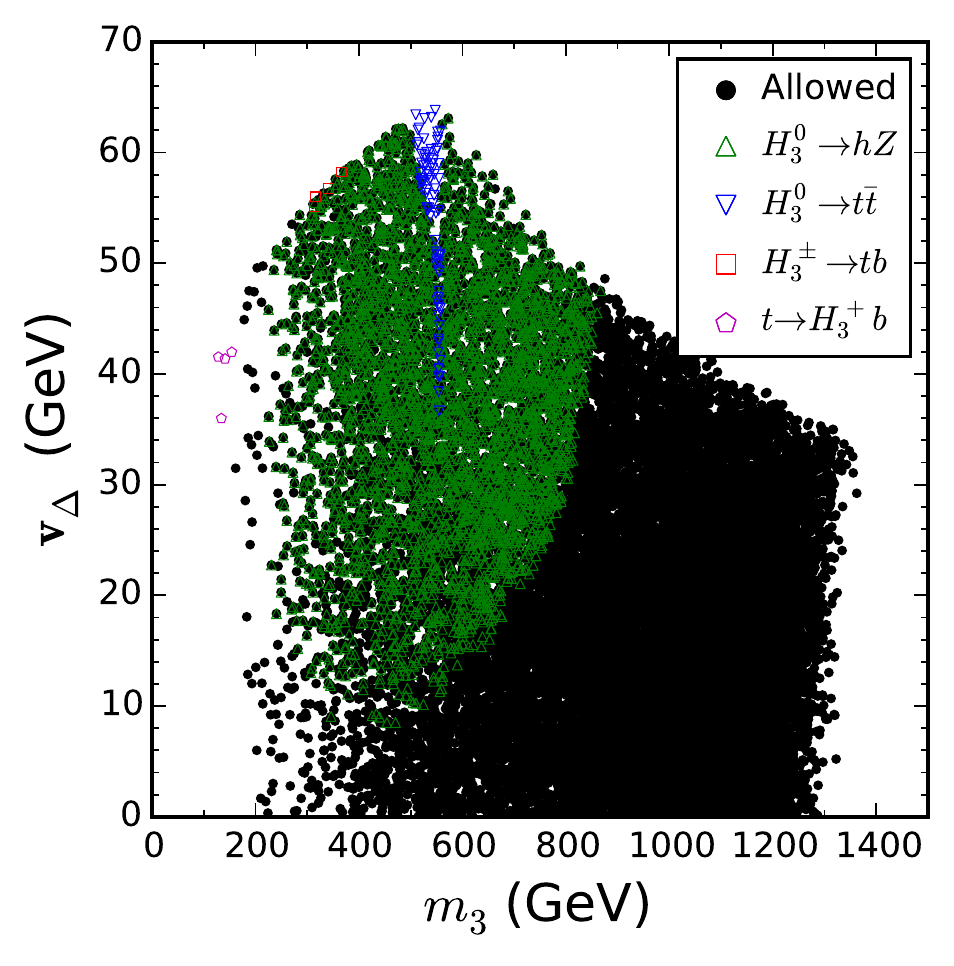}~
    \includegraphics[width=0.465\linewidth]{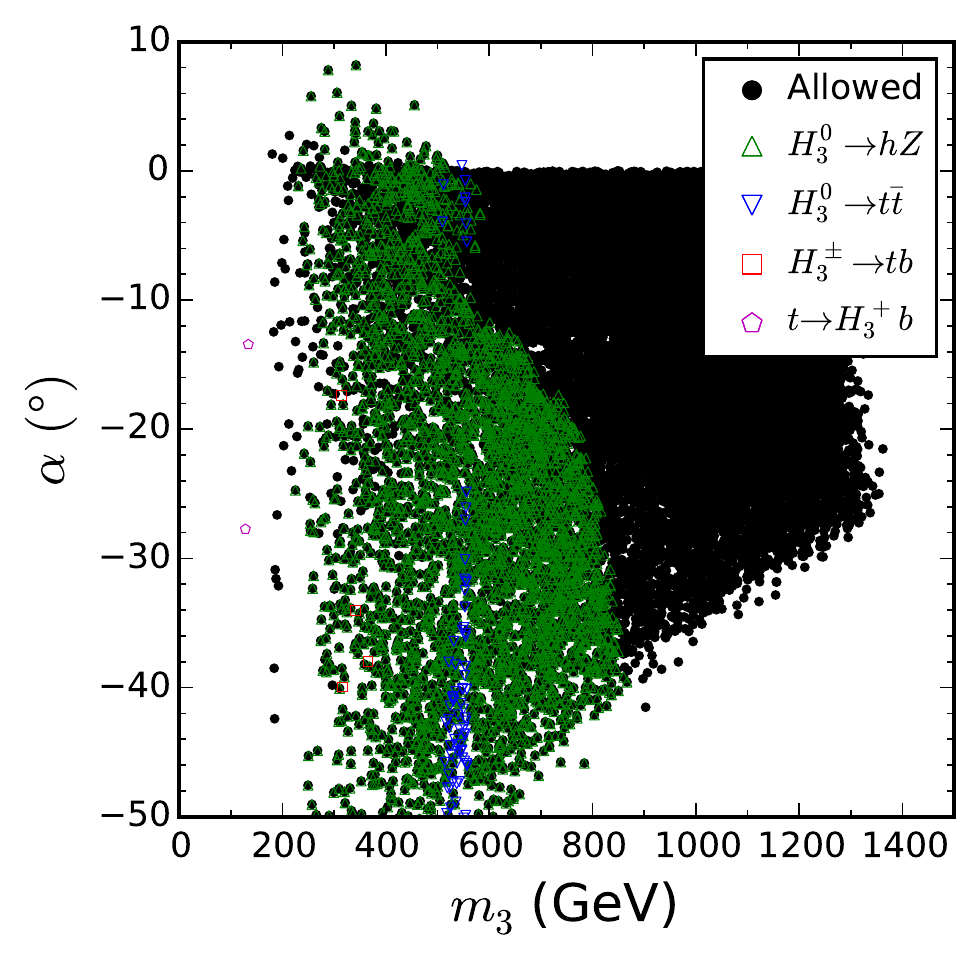}
	\caption{Distribution of survived points shown in the $v_\Delta-m_3$ plane (left) and $\alpha-m_3$ plane (right). The green, blue, red, and purple points are excluded by the $H_3^0\to hZ$, $H_3^0\to t\bar{t}$, $H^\pm\to tb$, and $t\to H_3^+ b$ searches respectively.}
	\label{fig:H3}
\end{figure}

The signature of a neutral triplet Higgs boson $H_3^0$ has been considered in Ref.~\cite{Chiang:2015kka}. Without direct couplings to gauge bosons, the promising signature is $gg\to H_3^0\to Zh$ for $m_Z+m_h<m_3<2m_t$, or $gg\to H_3^0 \to t\bar{t}$ for $m_3>2m_t$~\cite{Chiang:2015kka}. The charged triplet Higgs boson $H_3^+$ can decay into $\tau^+\nu$ for $m_3<m_t$ or into $t\bar{b}$ for $m_3>m_t+m_b$~\cite{Chiang:2012cn}. We therefore consider the following direct searches:
\begin{itemize}
	\item Search for a CP-odd Higgs boson $H_3^0$ decaying to $hZ$ \cite{Aad:2015wra,TheATLAScollaboration:2016loc}.
    \item Search for a heavy Higgs boson $H_3^0$ decaying to a top quark pair \cite{Aaboud:2017hnm}.
	\item Search for charged Higgs bosons decaying via $t\to H_3^+(\to \tau^+\nu)b$ or $H_3^\pm\to tb$ \cite{Aad:2014kga,Khachatryan:2015qxa}.
\end{itemize}
Our results are shown in Fig.~\ref{fig:H3} in the $v_\Delta-m_3$ and $\alpha-m_3$ planes. Among the constraints from those searches, $H_3^0\to hZ$ sets the most stringent one. In particular, in the mass region $200\lesssim m_3\lesssim 500~\GeV$, where BR($H_3^0\to hZ$) is dominant or relatively large, $v_\Delta$ can be pushed down as low as $10~\GeV$ under certain circumstances.

\subsection{Quintuplet Searches}

Being independent of the singlet mixing angle $\alpha$, the constraints on the quintuplet scalars are only sensitive to the VEV $v_\Delta$ and their mass $m_5$. In Ref.~\cite{Chiang:2015kka} the constraint on $v_\Delta$ has been obtained as a function of the exotic Higgs boson mass via various channels for the additional neutral scalars in the GM model. In this paper, we adopt the constraints from the decay channels $H_5^0\to\gamma\gamma$ and $H_5^0\to ZZ$ through the VBF mechanism.

In Refs.~\cite{Aad:2015nfa,CMS:2016szz} a search was performed for heavy charged scalars decaying to $W^\pm$ and $Z$ bosons at $\sqrt{s}=13~\TeV$~LHC. The upper limits on the cross section for the production of charged Higgs bosons times branching fractions to $W^\pm Z$ are transformed to the exclusion limits on $v_\Delta$ versus $m_5$ in the GM model.

\begin{figure}
	\centering
	\includegraphics[width=0.45\linewidth]{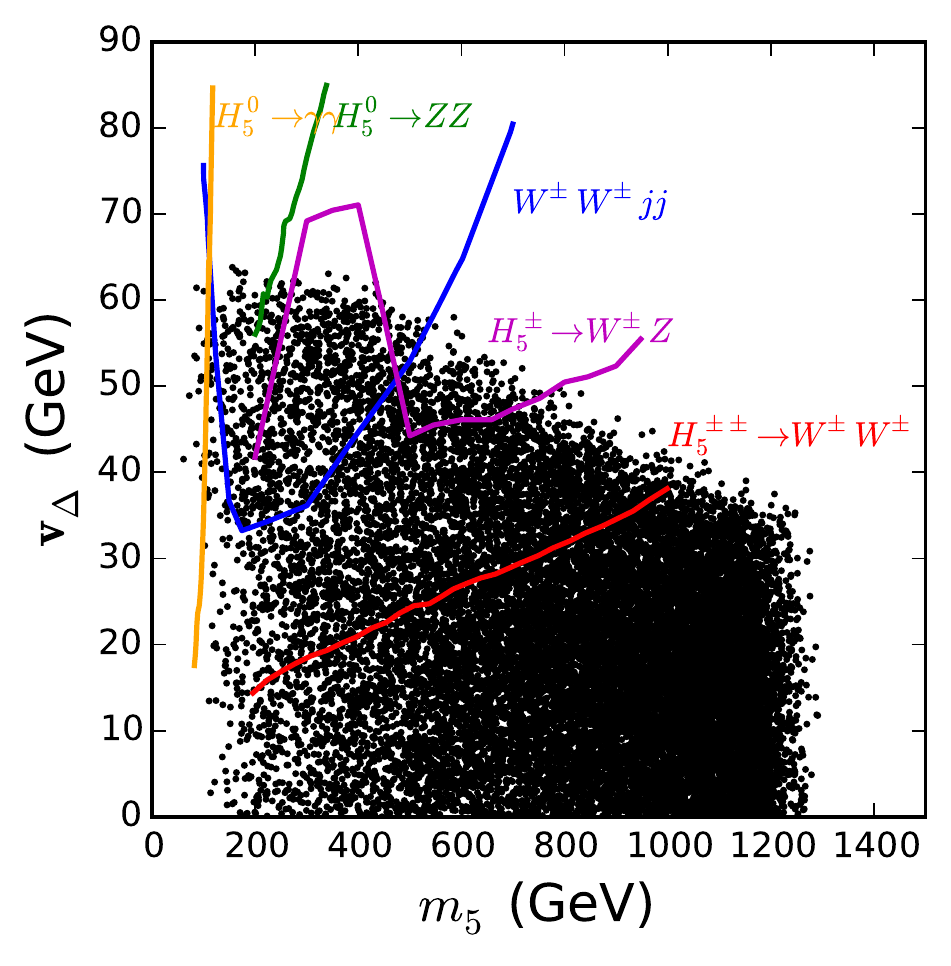}
	\caption[H5]{Constraints from searches for quintuplet Higgs bosons on the $v_\Delta-m_5$ plane. Points above exclusion curves are eliminated.}
	\label{fig:h5}
\end{figure}

The experimental constraints on the $H_5^{++}$ mass were studied in the GM model in Ref.~\cite{Chiang:2014bia} by recasting the ATLAS measurement of the cross section for the like-sign diboson process $pp\to W^\pm W^\pm jj$. The $W^+W^+W^-W^-$ vertex is effectively modified by mediations of the doubly-charged Higgs bosons $H^{\pm\pm}$. That the relevant $W^\pm W^\pm H^{\mp\mp}$ vertex is proportional to $v_\Delta$ can be used to exclude parameter space on the plane of $v_\Delta$ and $m_5$. In this work we also take into account the latest search for like-sign $W$ boson pairs by the CMS~\cite{Sirunyan:2017ret}.

Additional subsidiary constraints are as follows:
\begin{itemize}
	\item An absolute lower bound on the doubly-charged Higgs mass from the ATLAS like-sign dimuon data was obtained in Ref.~\cite{Kanemura:2014ipa}, which gives $m_5\gtrsim 76$ GeV.
	\item An upper bound on $s_H$ for $m_5\sim 76-100~\GeV$ can be obtained using the results of a decay-model-independent search for new scalars produced in association with a $Z$ boson from the OPAL detector at the LEP collider \cite{Abbiendi:2002qp}.
	\item A limit $\tan \theta_H < 10/3$ is imposed to avoid a nonperturbative top quark Yukawa coupling~\cite{Barger:1989fj}.
\end{itemize}

In Fig.~\ref{fig:h5}, we summarize the constraints from searches for quintuplet Higgs bosons on the $v_\Delta-m_5$ plane. In the low mass region $76~\GeV<m_5<110~\GeV$, the most stringent constraint comes from the neutral Higgs boson decay $H_5^0\to \gamma\gamma$ through the VBF production. In the mass interval $110-200~\GeV$, the like-sign diboson process $pp\to W^\pm W^\pm jj$ via the doubly-charged Higgs boson $H_5^{\pm\pm}$ gives the best bound. Above $200~\GeV$, the most severe constraint is set by the latest search for like-sign $W$ boson pairs, which could exclude $v_\Delta\gtrsim 15~\GeV$ when $m_5\sim 200~\GeV$.

\subsection{Higgs Signal Strengths}

\begin{figure}[!htbp]
	\centering
	\includegraphics[width=0.45\linewidth]{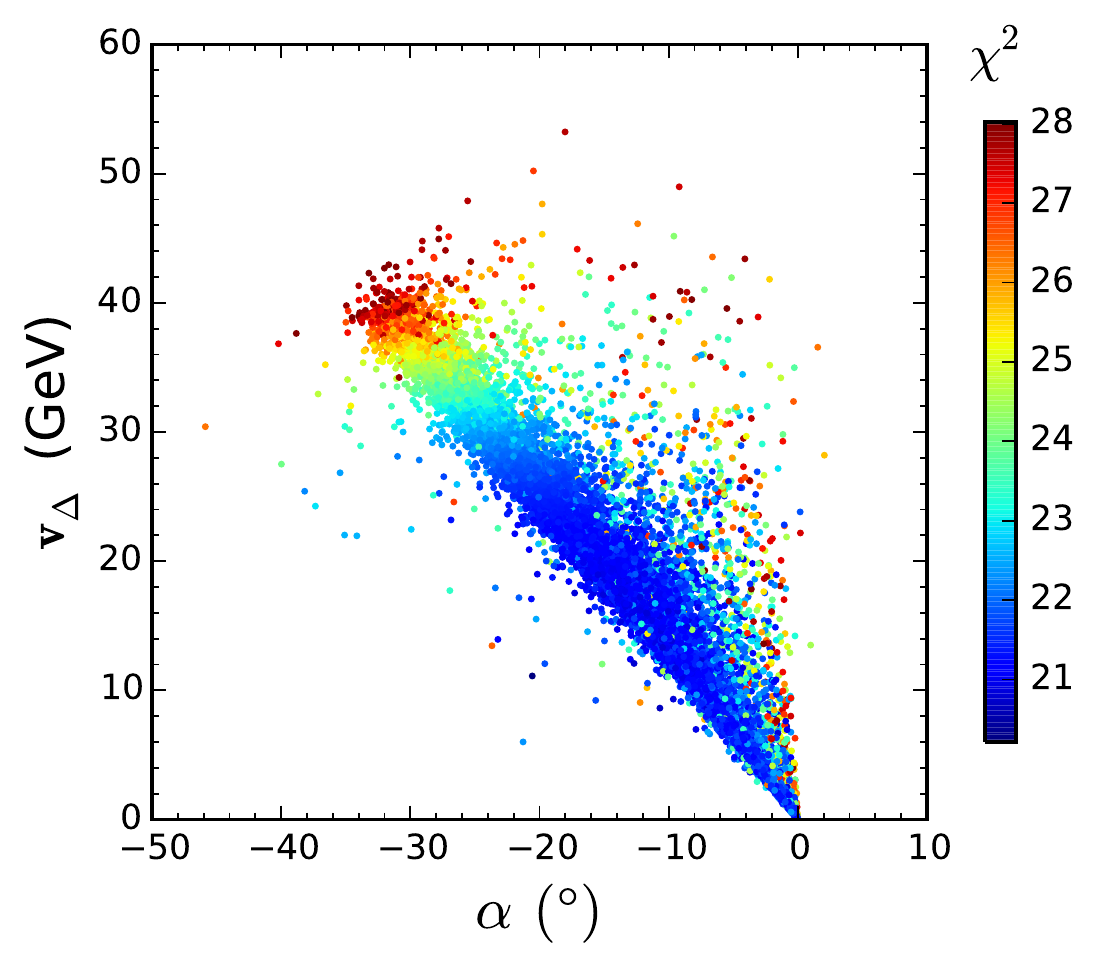}
	\caption[H5]{Scatter plot of the Higgs signal strength $\chi^2$ fit with a $2\sigma$ range shown in the $\alpha-v_\Delta$ plane.}
	\label{fig:chi}
\end{figure}

The signal strengths of the SM-like Higgs boson production and decay in various channels can provide significant constraints on its couplings to the SM particles in the GM model~\cite{Belanger:2013xza}. The signal strength for a specific production and decay channel $i\to h\to f$ is defined as
\begin{equation}
\begin{aligned}
\mu_{i}^f
=\frac{\sigma_i\times \text{BR}_f}{\sigma^{\text{SM}}_i\times \text{BR}^{\text{SM}}_f},
\end{aligned}
\end{equation}
where $\sigma_i~(\sigma^{\text{SM}}_i)$ is the reference value (SM prediction) of the Higgs production cross section for $i\to h$, and $\text{BR}_f~(\text{BR}^{\text{SM}}_f)$ the branching ratio for the decay $h\to f$. We include the production channels via the gluon fusion ($ggF$), the vector boson fusion (VBF), the associated production with a vector boson ($Vh$) and with a pair of top quarks ($tth$), and the decay channels $h\to\gamma\gamma,~ZZ,~W^\pm W^\mp,~\tau^\pm\tau^\mp,~b\bar{b}$. With the experimental values of $\mu_X^{\text{exp}}$ and standard deviation $\Delta \mu_X^{\text{exp}}$~\cite{Khachatryan:2016vau}, we build a $\chi^2$ value for each allowed point as
\begin{equation}
\chi^2=\sum_X\left(\frac{\mu_X^{\text{exp}}-\mu_X}{\Delta \mu_X^{\text{exp}}}\right)^2,
\end{equation}
where the sum extends over all channels mentioned above.

From Eqs.~(\ref{eqn:hvv}, \ref{eqn:khtt}), we are aware that the SM-like Higgs couplings involved in the signal strengths depend only on the triplet VEV ($v_\Delta$) and the singlet mixing angle ($\alpha$). Therefore the constraints on $v_\Delta$ and $\alpha$ can be directly extracted from data without specifying other parameters. In Fig.~\ref{fig:chi} we show the scatter plot of $\chi^2$ values on the $\alpha-v_\Delta$ plane within a $2\sigma$ range. It is clear that the measurement of the Higgs signal strengths is most sensitive to the region with large $v_\Delta$ and large $|\alpha|$, where large deviations of Higgs couplings from the SM take place. Hence the large $\chi^2$ region, e.g., $v_\Delta\sim50$ GeV and $\alpha\sim-40^\circ$ would be excluded by future operations of LHC~\cite{Chiang:2015amq}.

\subsection{Future Experimental Constraints}

For completeness we also presented in Fig.~\ref{fig:constraints} the constraints of the Higgs signal strength $\chi^2$ fit on the $\alpha-v_\Delta$ plane based on the projection results from 14~TeV LHC with an integrated luminosity of $300~\fb^{-1}$ (LHC@300) and $3000~\fb^{-1}$ (HL-LHC@3000)~\cite{ATLAS:HL-LHC} and from ILC with an integrated luminosity of $1150~\fb^{-1}$ at 250~GeV (ILC250) and $1600~\fb^{-1}$ at 500~GeV (ILC500)~\cite{Asner:2013psa}. The LHC (HL-LHC) result is performed on the ATLAS detector with only statistical and experimental systematic uncertainties taken into account. The expected precision is given as the relative uncertainty in the signal strength with the central values all assumed to be unity. In principle, this assumption applies only to the SM, but we employ these anticipated results as a reference so that we could be clear to what extent they can impose a constraint on the parameter space. It is clear from Fig.~\ref{fig:constraints} that the constrains from LHC@300, HL-LHC@300, ILC250, and ILC500 gradually become more and more stringent. Basically speaking, the deviations in Higgs couplings are determined by $\alpha$ and $v_\Delta$. Our fitting results show that LHC@300 (HL-LHC@3000) could approximately  exclude $v_\Delta\gtrsim 30~(20)~\GeV$ while ILC250 and ILC500 could further push it down to about $v_\Delta\gtrsim 10~\GeV$.

Let us consider the scale factors $\kappa_{hVV}$ and $\kappa_{hf\bar{f}}$ as an example under the assumption that no obvious deviations in Higgs couplings from SM values will be observed. At LHC@300, $\kappa_{hVV}$ and $\kappa_{hf\bar{f}}$ could be constrained within the ranges $[0.92,1.08]$ and $[0.93,1.13]$ respectively, while at LHC@3000 they could be further narrowed down to $\kappa_{hVV}\in[0.97,1.04]$ and $\kappa_{hf\bar{f}}\in[0.93,1.06]$. At $e^+e^-$ colliders, e.g., ILC, precision measurements of the Higgs productions and decays can rigorously constrain the parameter space. At ILC250 where $e^+e^-\to hZ$ dominates, measurements of the Higgs signal strength in this channel could give $\kappa_{hVV}\in[0.998,1.02]$ and $\kappa_{hf\bar{f}}\in[0.995,1.007]$. And at ILC500 measurements via the $e^+e^-\to hZ$ and $e^+e^-\to h\nu_e\bar{\nu}_e$ channels would yield $\kappa_{hVV}\in[0.998,1.01]$ and $\kappa_{hf\bar{f}}\in[0.998,1.004]$. On the other hand, if the on-going LHC observes a certain hint of Higgs coupling deviation , for instance, $\kappa_{hVV}>1$ for the most optimistic case, the future operation of ILC will be hopeful to confirm it. In this way, the GM model would be strongly favored by the deviation $\kappa_{hVV}>1$.

\section{Higgs Production at $e^+e^-$ Colliders}
\label{sec:phenomenology}

\begin{figure}[!htbp]
	\centering
	\includegraphics[width=0.45\linewidth]{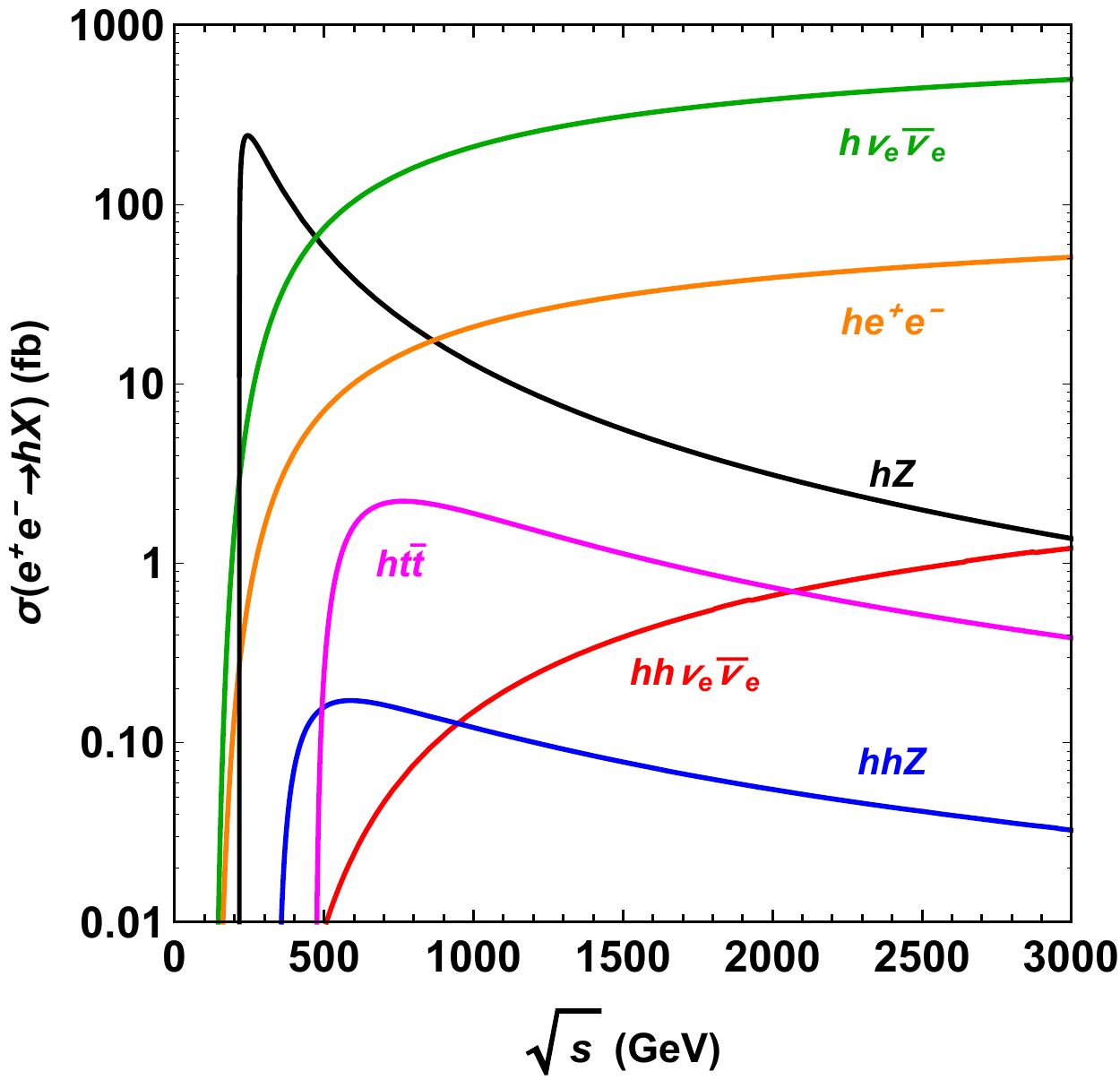}
	\caption[H5]{Cross section as a function of CM energy for dominant Higgs production processes in the SM at $e^+e^-$ colliders with $m_h=125$ GeV.}
	\label{fig:cs_ee_hX}
\end{figure}

In this section  we will study the dominant single and double SM-like Higgs production at future $e^+e^-$ colliders in the GM model. For comparison we first reproduce in Fig.~\ref{fig:cs_ee_hX} the dominant Higgs production cross sections in the SM. The Higgs-strahlung (HS) process ($e^+e^-\to hZ$) is dominant for a CM energy $\sqrt{s}<500~\GeV$. At higher energy, the Higgs production is dominated by the $WW$ fusion process ($e^+e^-\to h\nu_e\bar{\nu}_e$), while the $ZZ$ fusion process ($e^+e^-\to he^+e^-$) also becomes significant. The subdominant processes such as $e^+e^-\to ht\bar{t}$, $e^+e^-\to hhZ$ and $e^+e^-\to hh \nu_e\bar{\nu}_e$ provide access to the top Yukawa coupling and the Higgs trilinear self-coupling.

\begin{figure}[!htbp]
	\centering
	\includegraphics[width=0.9\linewidth]{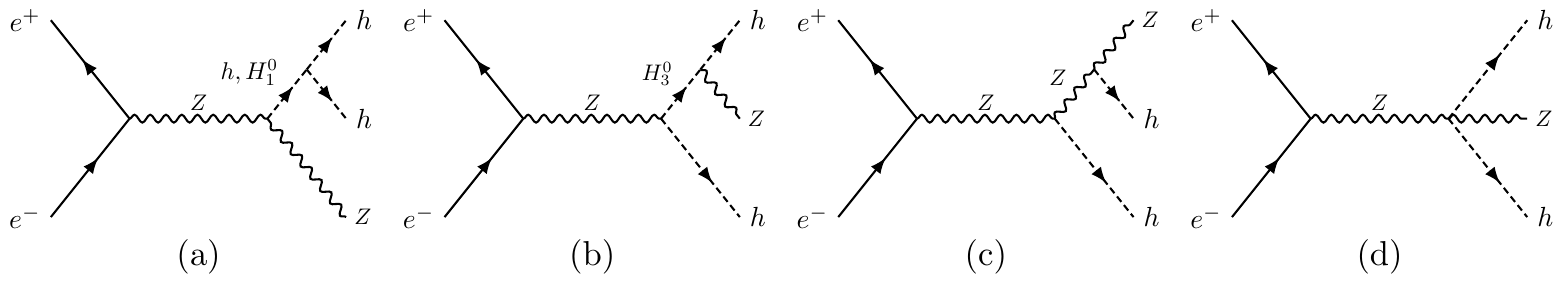}\\
	\caption{Feynman diagrams for $hhZ$ production at $e^+e^-$ colliders.}
	\label{fig:feyn-hhz}
\end{figure}

It is clear that due to the $s$-channel topology (see Figs.~\ref{fig:feyn-single}, \ref{fig:feyn-htt}, \ref{fig:feyn-hhz}), the $hZ$, $ht\bar{t}$, and $hhZ$ production cross sections become maximal near the thresholds and decrease as collision energy goes up. On the contrary, the VBF ($h\nu_e\bar{\nu}_e$, $he^+e^-$, and $hh \nu_e\bar{\nu}_e$) cross sections increase as $\ln\sqrt{s}$ due to their $t$-channel topology (see Figs.~\ref{fig:feyn-single}, \ref{fig:feyn-hhvv}). These dominant processes can be divided into three types according to the couplings involved, namely the Higgs couplings to gauge bosons (Fig.~\ref{fig:feyn-single}) and to the top quark (Fig.~\ref{fig:feyn-htt}), and the trilinear Higgs self-couplings (Figs.~\ref{fig:feyn-hhz}, \ref{fig:feyn-hhvv}), which we now investigate in detail for the GM model.

\subsection{Production via Higgs-strahlung and vector boson fusion}

\begin{figure}[!htbp]
	\centering
	\includegraphics[width=0.9\linewidth]{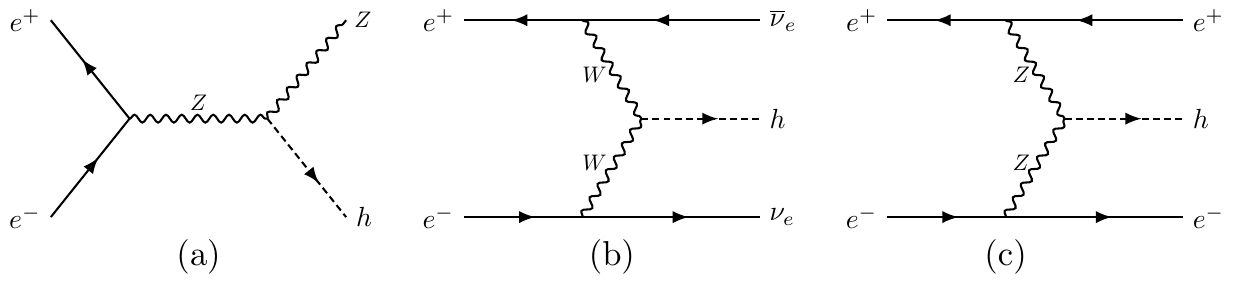}\\
	\caption{Feynman diagrams for Higgs production via HS (a) and VBF (b, c) at $e^+e^-$ colliders.}
	\label{fig:feyn-single}
\end{figure}

In Fig.~\ref{fig:feyn-single} we depict the Feynman diagrams for single Higgs production at $e^+e^-$ colliders that involves only Higgs couplings to weak gauge bosons. The amplitudes for both HS and VBF processes are modified in the GM model by the same ratio of the Higgs-gauge couplings from the SM,
\begin{equation}
\label{eqn:hvv}
\kappa_{hVV}=\frac{g_{hVV}}{g_{hVV}^{\text{SM}}}
= c_\alpha c_H-\frac{2\sqrt{6}}{3}s_\alpha s_H,
\end{equation}
where $c_H$ and $s_H$ are defined in Eq.~(\ref{eqn:chsh}) in terms of $v_\Delta$. We can thus extract $\kappa_{hVV}^2$ by measuring these cross sections and set constraints on the parameters $v_\Delta$ and $\alpha$,

\begin{figure}[!htbp]
	\centering
	\includegraphics[width=0.45\linewidth]{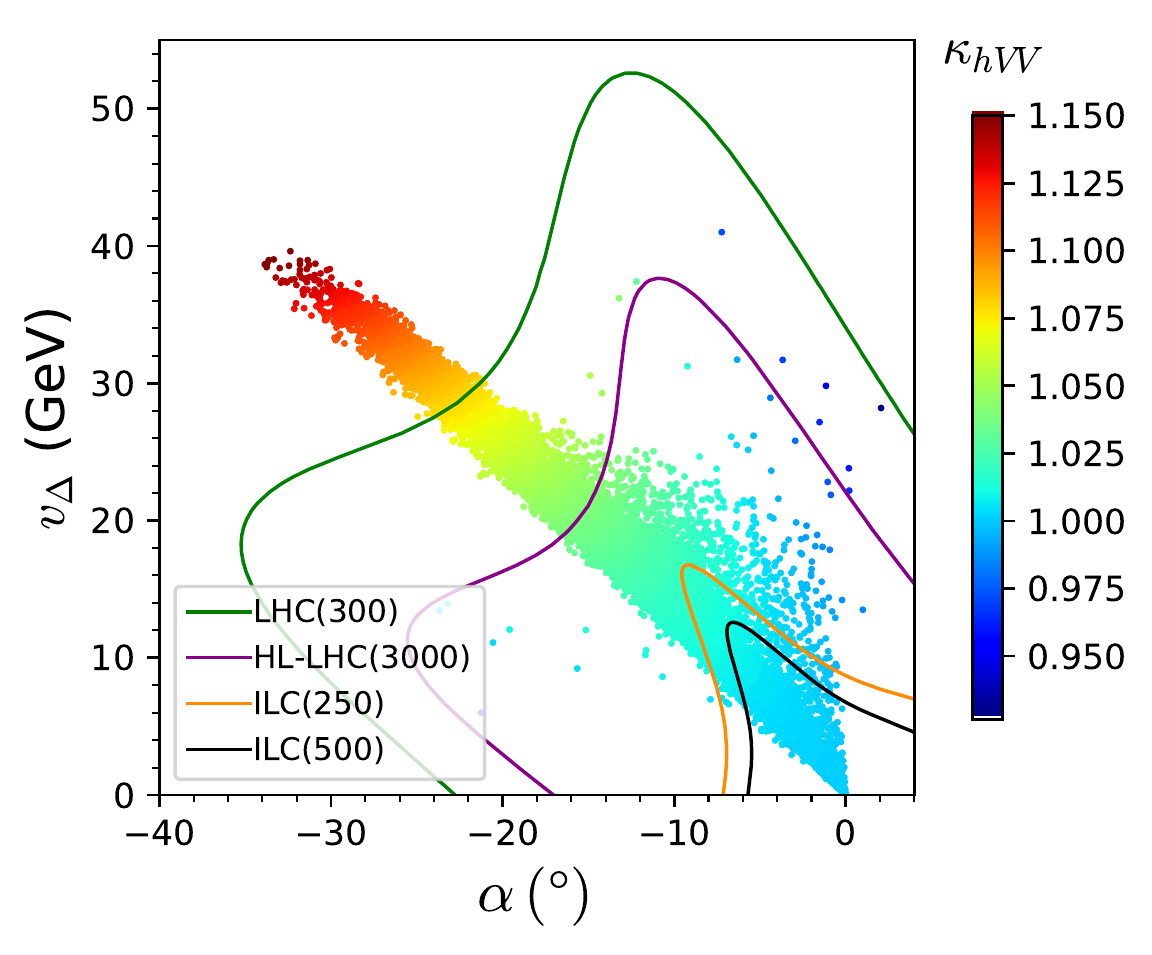}\\
	\caption{Predicted value of $\kappa_{hVV}$ in the $v_\Delta-\alpha$ plane after imposing all the constraints in Sec.~\ref{sec:constraints}.}
	\label{fig:khVV}
\end{figure}

The predicted value of $\kappa_{hVV}$ in the $v_\Delta-\alpha$ plane is shown in  Fig.~\ref{fig:khVV}. It is obvious that an $\mathcal{O}(10\%)$ deviation of $\kappa_{hVV}$ from unity is still viable. Although the allowed $\kappa_{hVV}$ is in a range of about 0.93-1.15, most of the survived points tend to have $\kappa_{hVV}>1$.  The LHC@300 will mostly exclude $\kappa_{hVV}\gtrsim1.1$, while HL-LHC@3000 could probe down to $\kappa_{hVV}\gtrsim1.05$. The scale factor $\kappa_{hVV}$ is expected to be measured at future $e^+e^-$ colliders with high precision. For example, the measurements for the $hVV$ couplings may reach an accuracy of 1\% at CEPC (250~GeV, 5 ab$^{-1}$) and ILC (500~GeV, 500~fb$^{-1}$)~\cite{CEPC-SPPCStudyGroup:2015esa,Asner:2013psa}. Hence, the GM model could be probed indirectly if a large enough deviation of $\kappa_{hVV}$ from unity is measured. Especially, a measured $\kappa_{hVV}>1$ would be strong evidence in favor of the GM model. If $\kappa_{hVV}$ turns out to be consistent with unity, a precise measurement of it would put a stringent bound on the GM model parameter space.

The cross section for the HS process is~\cite{Barger:1993wt}
\begin{equation}
\sigma(e^+e^-\to Zh)=\frac{G_F^2m_Z^4}{96\pi s}(V_e^2+A_e^2)
\frac{\sqrt{\beta}(\beta+12 r_Z)}{(1-r_Z)^2}\kappa_{hVV}^2,
\end{equation}
where the $Z$ couplings to the fermion $f$ of electric charge $Q_f$ are $V_f=2I_3^f-4Q_fs_W^2,~A_f=2I_3^f$, with $I_3^f=\pm1/2$ being the third weak isospin of the left-handed fermion $f$, and $s_W^2=\sin^2\theta_W$ with $\theta_W$ being the Weinberg angle. And $\beta=(1-r_Z-r_h)^2-4r_Zr_h$ is the usual two-body phase space function with $r_{h,Z,W}\equiv m_{h,Z,W}^2/s$, etc.

\begin{figure}[!htbp]
	\centering
   \includegraphics[width=0.45\linewidth]{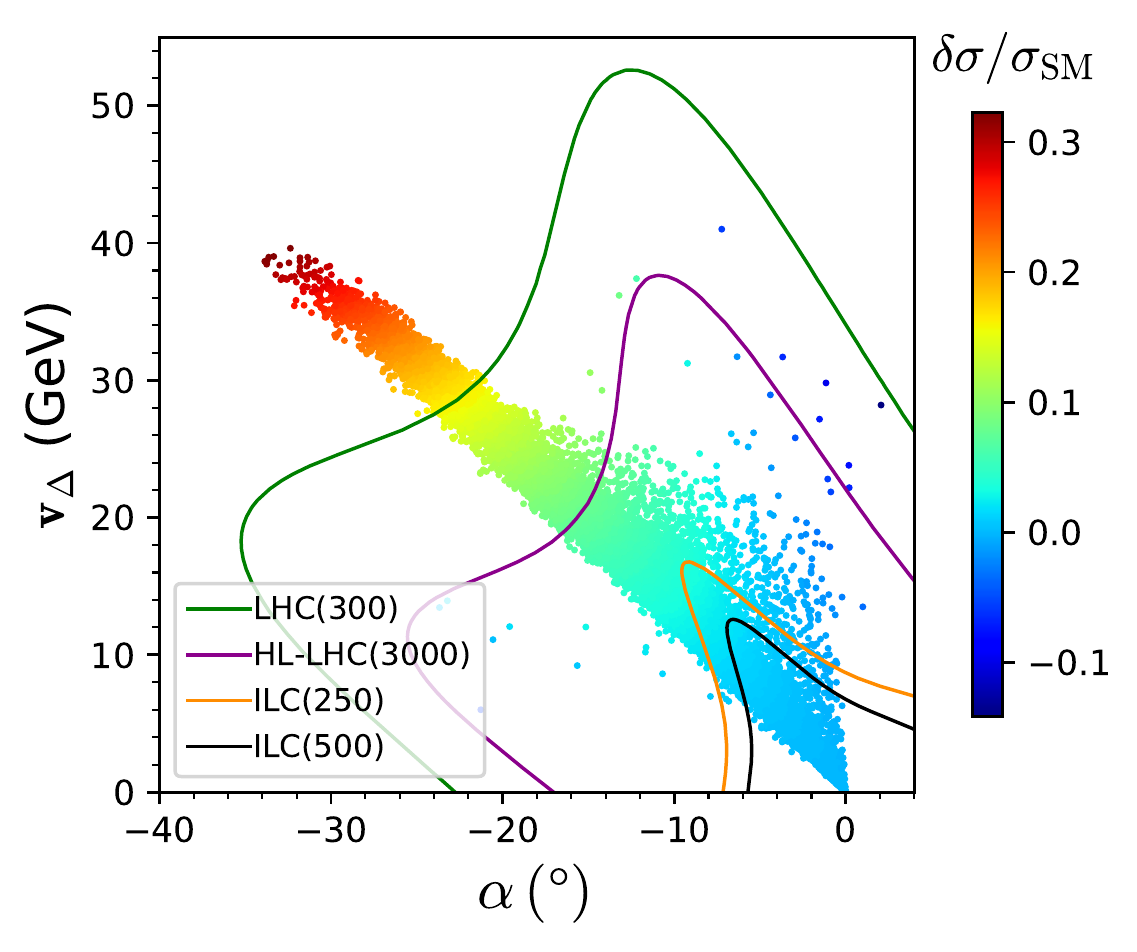}
	\caption{Distribution of relative corrections $\delta\sigma/\sigma_{\text{SM}}$ in the $v_\Delta-\alpha$ plane for HS and VBF processes.}
    \label{fig:hZ-hvv}
\end{figure}

The total cross section for the $WW~(ZZ)$ fusion process is~\cite{Djouadi:1996uj}
\begin{equation}
\sigma_{VV}=\frac{G_F^3m_V^4}{64\sqrt{2}\pi^3}\int_{r_h}^{1}dx\int_{x}^{1}
dy\frac{(V_V^2+A_V^2)^2f(x,y)+4V_V^2A_V^2g(x,y)}{[1+(y-x)/r_V]^2}\kappa_{hVV}^2,
\label{eq:sigmaVV}
\end{equation}
where $V$ denotes either $W$ or $Z$, $V_W=A_W=\sqrt{2}~(V_Z=V_e,~A_Z=A_e)$ for the $WW~(ZZ)$ fusion respectively, and
\begin{align}
f(x,y)&=\left(\frac{2x}{y^3}-\frac{1+2x}{y^2}+\frac{2+x}{2y}-\frac{1}{2}\right)
\left[\frac{z}{1+z}-\ln(1+z)\right]+\frac{x}{y^3}\frac{z^2(1-y)}{1+z},\\
g(x,y)&=\left(-\frac{x}{y^2}+\frac{2+x}{2y}-\frac{1}{2}\right)\left[\frac{z}{1+z}-\ln(1+z)\right],
\end{align}
with $z=y(x-r_h)/(xr_V)$.

The HS and VBF processes are within the reach of all future $e^+e^-$ colliders mentioned in Sec.~\ref{sec:introduction}. Their cross sections are corrected by the same factor of $\kappa_{hVV}^2$, so that their relative corrections compared to the SM $\delta\sigma/\sigma_{\text{SM}}=\kappa_{hVV}^2-1$ are the same and independent of the collision energy. From Fig.~\ref{fig:hZ-hvv}, we see that the cross sections could maximally increase by 32\% or decrease by 13\% with the current constraints. Even if no clear deviation would be found at LHC@300 or HL-LHC@3000, an increase of up to 10\% in the HS and VBF cross section would still be possible at the ILC.

\subsection{Associated Production with Top Quarks}
\label{subsec:htt}
\begin{figure}[!htbp]
	\centering
	\includegraphics[width=0.9\linewidth]{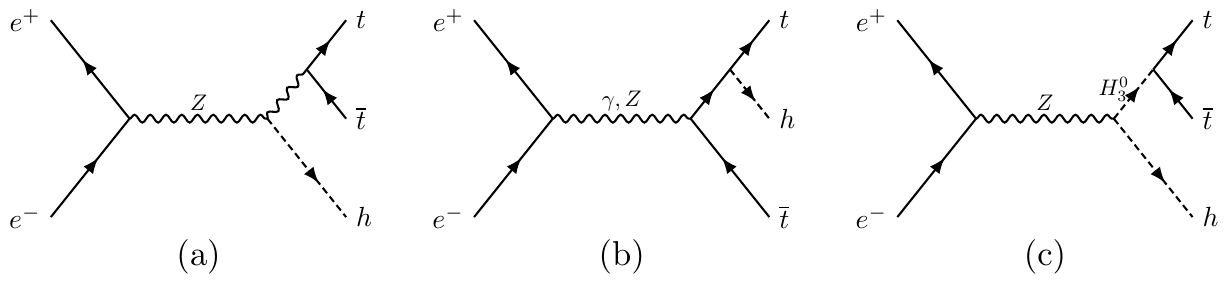}
\\
	\caption{Feynman diagrams for $ht\bar t$ production at $e^+e^-$ colliders.}
	\label{fig:feyn-htt}
\end{figure}

\begin{figure}
	\centering
    \includegraphics[width=0.32\linewidth]{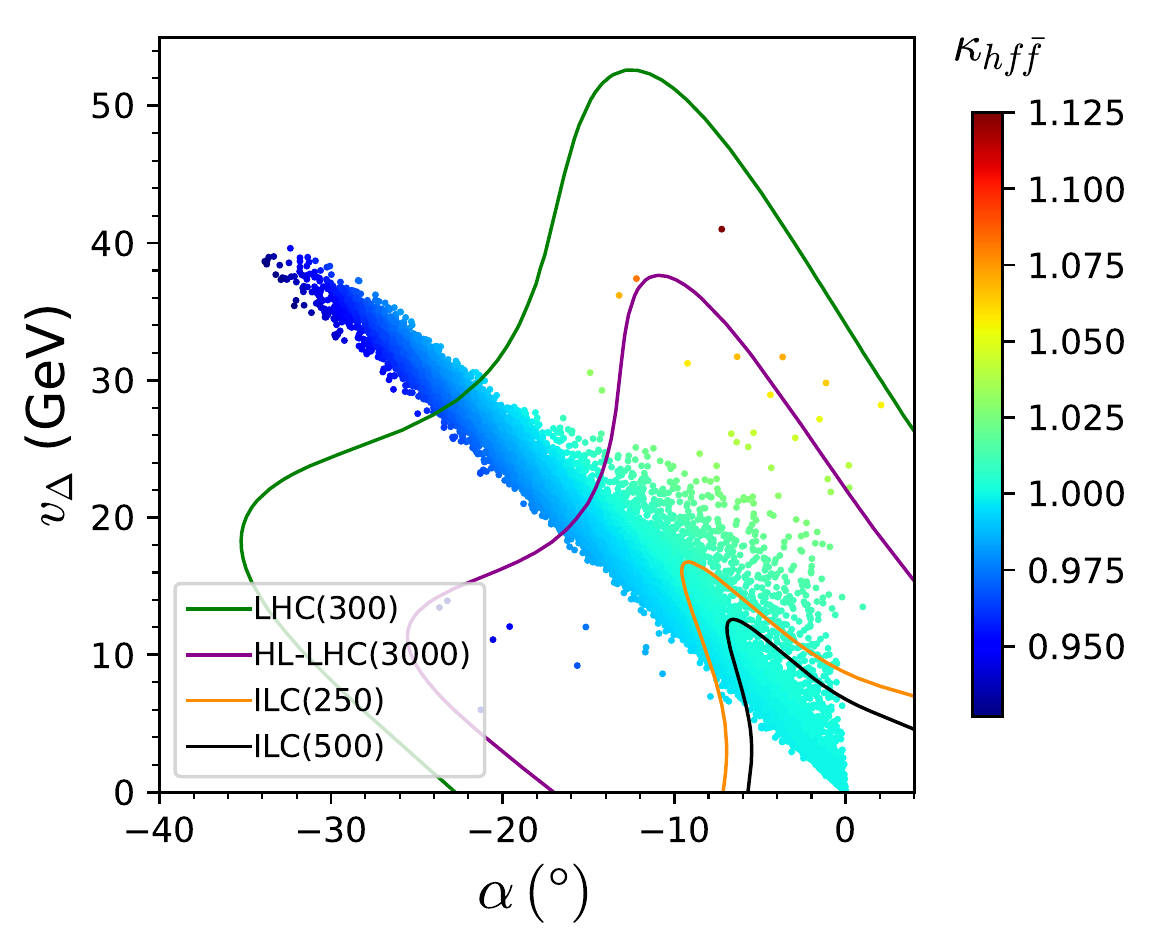}
	\includegraphics[width=0.31\linewidth]{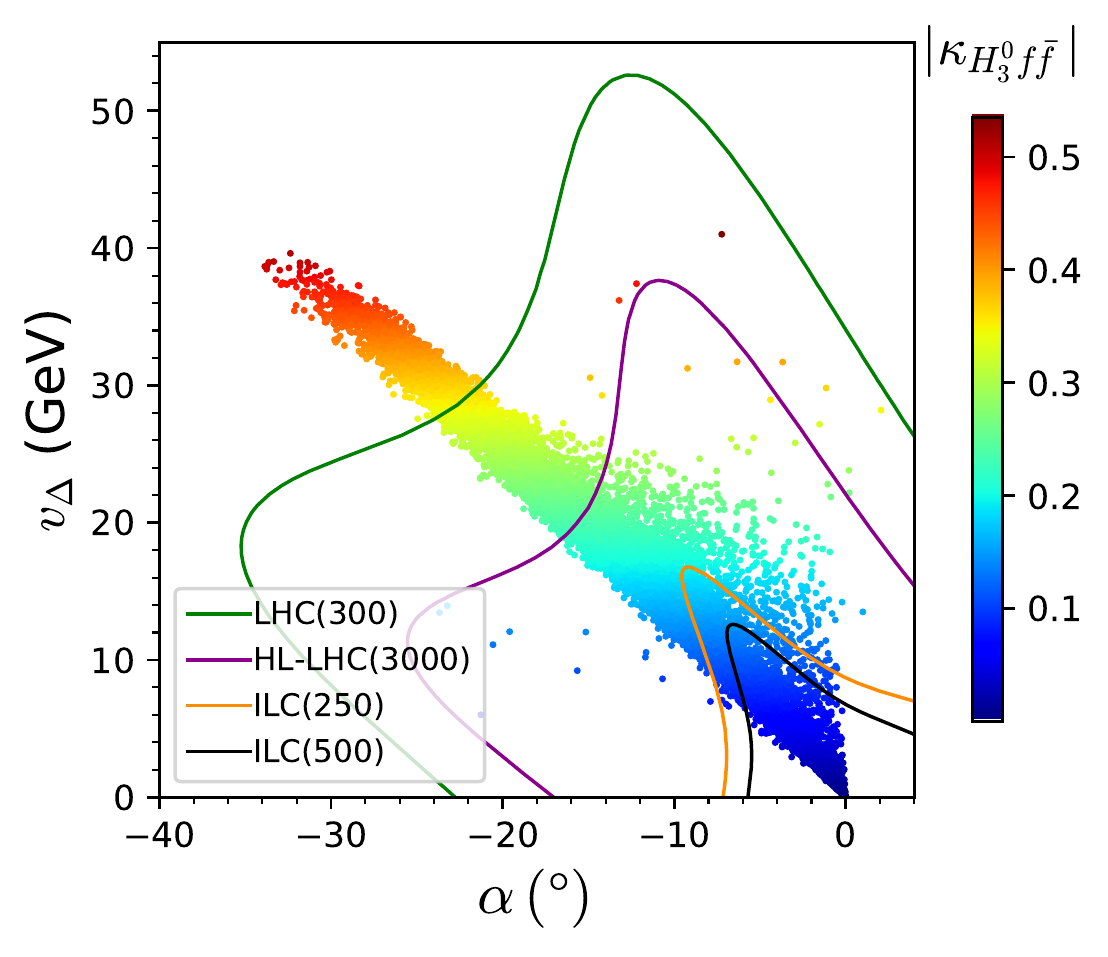}
    \includegraphics[width=0.32\linewidth]{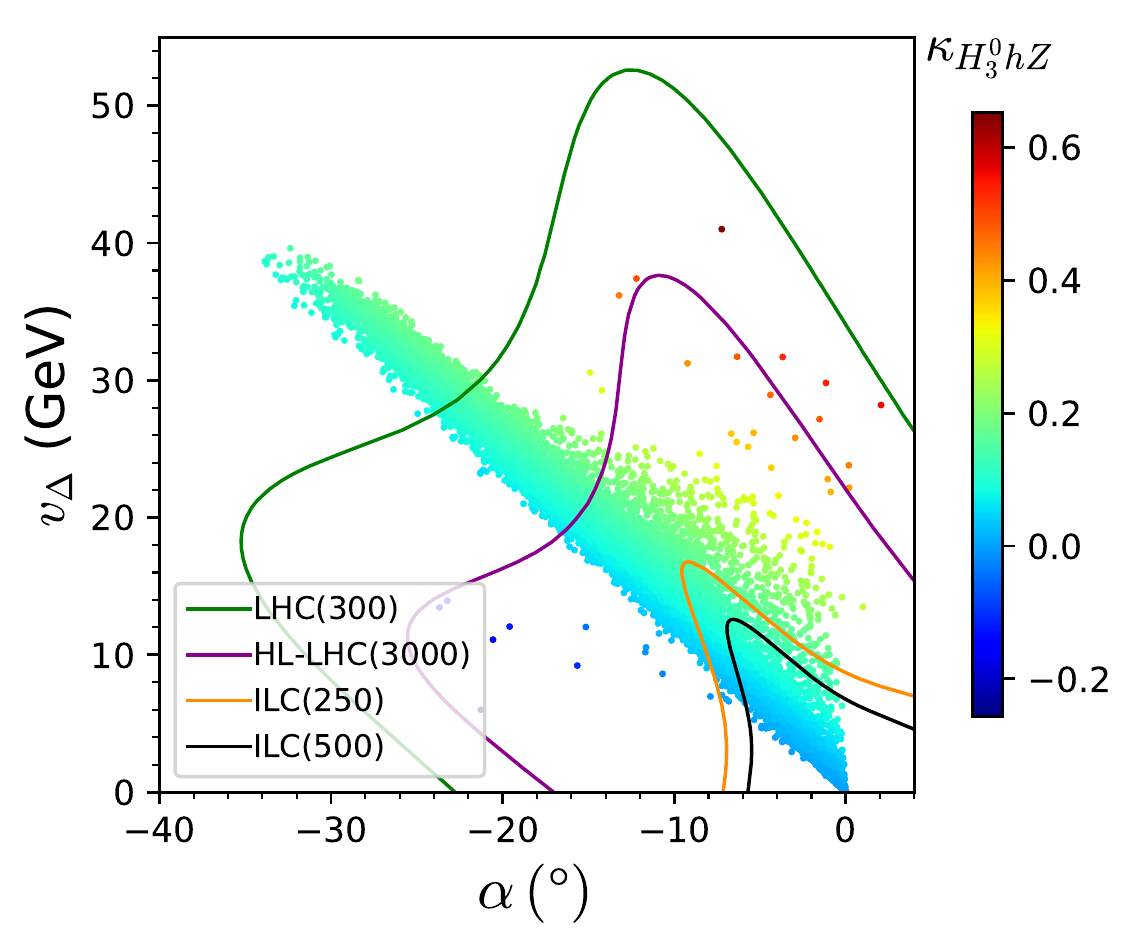}
	\caption{Predicted values of $\kappa_{hf\bar{f}}$, $|\kappa_{H_3^0f\bar{f}}|$, and $\kappa_{H_3^0hZ}$ in the $v_\Delta-\alpha$ plane after imposing all the constraints in Sec.~\ref{sec:constraints}.}
	\label{fig:khtt}
\end{figure}

The associated Higgs production with a top quark pair is an important process to measure the top quark Yukawa coupling at a linear collider~\cite{Gaemers:1978jr,Djouadi:1992gp}. Compared to the SM case, the existing interactions are modified and in addition there is a new contribution in the GM model that is mediated by the CP-odd heavy Higgs $H_3^0$, see Fig.~\ref{fig:feyn-htt}. The scale factor to the SM $h\bar f f$ coupling is, $\kappa_{hf\bar{f}}=c_\alpha/c_H$, and those involving the custodial triplet $H_3$ are
\begin{eqnarray}
\nonumber
&&-\kappa_{H_3^0 d\bar{d}}=\kappa_{H_3^0 u\bar{u}} = \tan\theta_H,\\
&&\kappa_{H_3^0hZ}=\kappa_{H_3^+hW^-}=c_\alpha s_H+\frac{2\sqrt{6}}{3}s_\alpha c_H,
\label{eqn:khtt}
\end{eqnarray}
which appear in the Feynman rules as follows:
\begin{eqnarray}\label{eq:definition}
\nonumber
H_3^0\bar ff&:&\kappa_{H_3^0 f\bar{f}}\frac{m_f}{v}\gamma_5,
\\
\nonumber
H_3^0hZ^\mu&:&\kappa_{H_3^0hZ}\frac{e}{2s_Wc_W}(p_h-p_3)_\mu,
\\
H_3^- hW^{+\mu}&:&-i\kappa_{H_3^+hW^-}\frac{e}{2s_W}(p_h-p_3)_\mu,
\end{eqnarray}
where $p_h~(p_3)$ is the incoming momentum of $h~(H_3^{0,+})$. The scanned results for these $\kappa$s are shown in Fig.~\ref{fig:khtt} in the $v_\Delta-\alpha$ plane. For the current constraints, we see that a deviation from unity as large as $\mathcal{O}(\pm10\%)$ is still allowed for $\kappa_{hf\bar{f}}$. The scale factor $\kappa_{H_3^0 f\bar{f}}$ does not depend on $\alpha$, and its magnitude can maximally reach about 0.54 and vanishes in the limit of $v_\Delta\to 0$, while $\kappa_{H_3^0hZ}$ lies in the interval of $-0.2$ to $0.6$. The future operation of LHC@300 will be capable of excluding points with $\kappa_{hf\bar{f}}\lesssim 0.975$ and $|\kappa_{H_3^0 f\bar{f}}|\gtrsim 0.35$, while ILC500 has the ability to constrain $\kappa_{hf\bar{f}}\approx 1$, $|\kappa_{H_3^0 f\bar{f}}|\lesssim 0.1$, and $\kappa_{H_3^0hZ}\in[0,0.2]$.

The cross section in the GM model can be adapted from those in the SM and MSSM~\cite{Djouadi:1992gp}
\begin{align}
\label{equ:htt}
\nonumber
\frac{d\sigma(e^+e^-\to ht\bar{t})}{dx_h}=&
N_c\frac{\sigma_0}{(4\pi)^2}\Bigg\{\Big[Q_e^2Q_t^2+
\frac{2Q_eQ_tV_eV_t}{16c_W^2s_W^2(1-r_Z)}
+\frac{(V_e^2+A_e^2)(V_t^2+A_t^2)}{256c_W^4s_W^4(1-r_Z)^2}\Big]G_1
\\
+&\frac{V_e^2+A_e^2}{256c_W^4s_W^4(1-r_Z)^2}
\Big[A_t^2\sum_{i=2}^{6}G_i+V_t^2(G_4+G_6)\Big]+
\frac{Q_eQ_tV_eV_t}{1-r_Z}G_6+G_7\Bigg\}.
\end{align}
Here $\sigma_0=4\pi\alpha_\text{QED}^2/(3s)$, $\alpha_\text{QED}$ is the fine structure constant, $N_c=3$, $x_h=2E_h/\sqrt{s}$ with $E_h$ the Higgs boson energy. The explicit expressions for the coefficients $G_i (i=1,\cdots 7)$ are given in Appendix~\ref{app:htt}.

\begin{figure}[!htbp]
	\centering
	\includegraphics[width=0.45\linewidth]{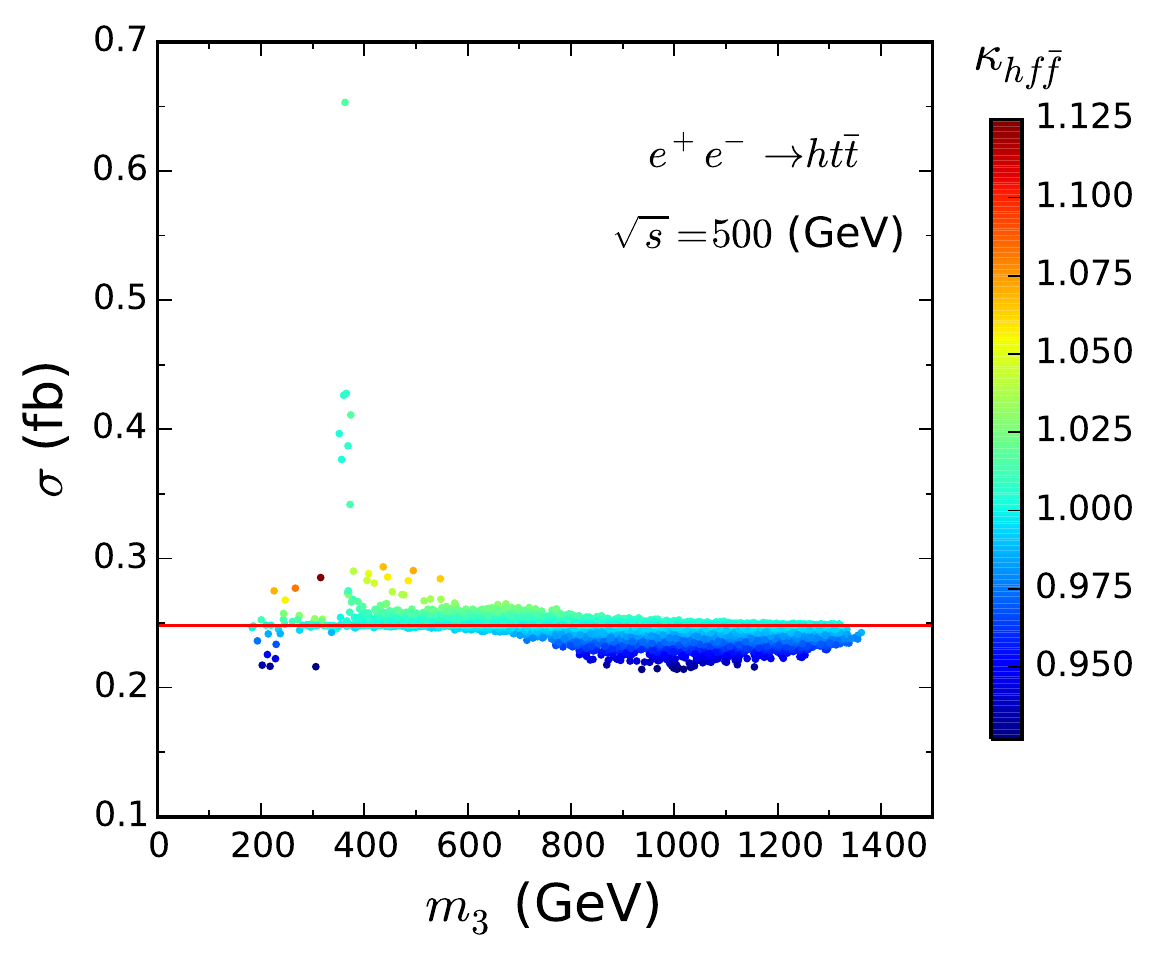}
	\includegraphics[width=0.44\linewidth]{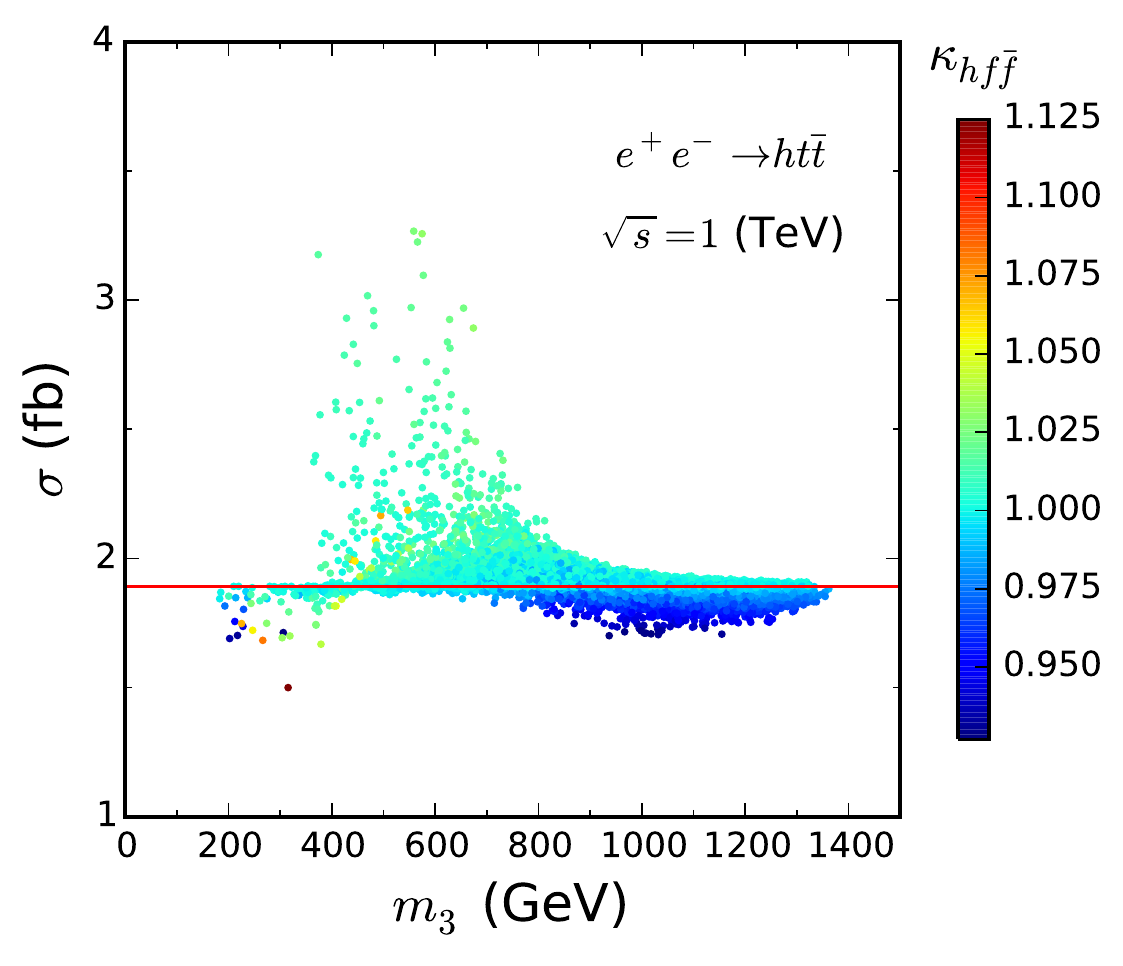}
	\caption{Cross section for $e^+e^-\to ht\bar{t}$ as a function of the $H_3^0$ mass $m_3$ at $\sqrt{s}=500~\GeV$ (left) and 1000~GeV (right). The horizontal line indicates the SM result.}
	\label{fig:htt}
\end{figure}

In Fig.~\ref{fig:htt} we present the cross section for $e^+e^-\to ht\bar{t}$ as a function of the $H_3^0$ mass $m_3$ at $\sqrt{s}=500,~1000~\GeV$. In the parameter region $2m_t<m_3<\sqrt{s}-m_h$, the subprocess $e^+e^-\to hH_3^0,~H_3^0\to t\bar{t}$ is resonant, where the cross section can be strongly enhanced by several times. When the mass of $H_3^0$ is far above the resonant region, the cross section reduces gradually to the SM value. Interestingly, in the resonant region with enhanced cross section,  we actually have $\kappa_{hf\bar{f}}\gtrsim1$ for most of the survived points, while above this region the cross section decreases with the decrease of $\kappa_{hf\bar{f}}$.  Combined with Fig.~\ref{fig:khtt}, most of points with $\kappa_{hf\bar{f}}<1$ will be excluded by LHC@300 and HL-LHC@3000. Hence, large decrease of the $ht\bar{t}$ production might not be possible. On the other hand, for $H_3^0$ resonantly produced, most of allowed points actually have $\kappa_{hf\bar{f}}\approx1$. Therefore, even if no large deviation in the Higgs coupling is observed at ILC250 or ILC500, significant enhancement of the $ht\bar{t}$ production may still be viable.  At CLIC the expected accuracy for cross section is about $8.4\%$ with an integrated luminosity of 1.5 ab$^{-1}$ at $\sqrt{s}=1.4~\TeV$~\cite{Abramowicz:2016zbo}, and at ILC the accuracy could reach $28\%~(6.3\%)$ at $500~(1000)~\GeV$~\cite{Asner:2013psa}. There is thus a good chance to test the $ht\bar{t}$ production at these high energy machines.

\subsection{Double Higgs Production}

\begin{figure}
	\centering
    \includegraphics[width=0.45\linewidth]{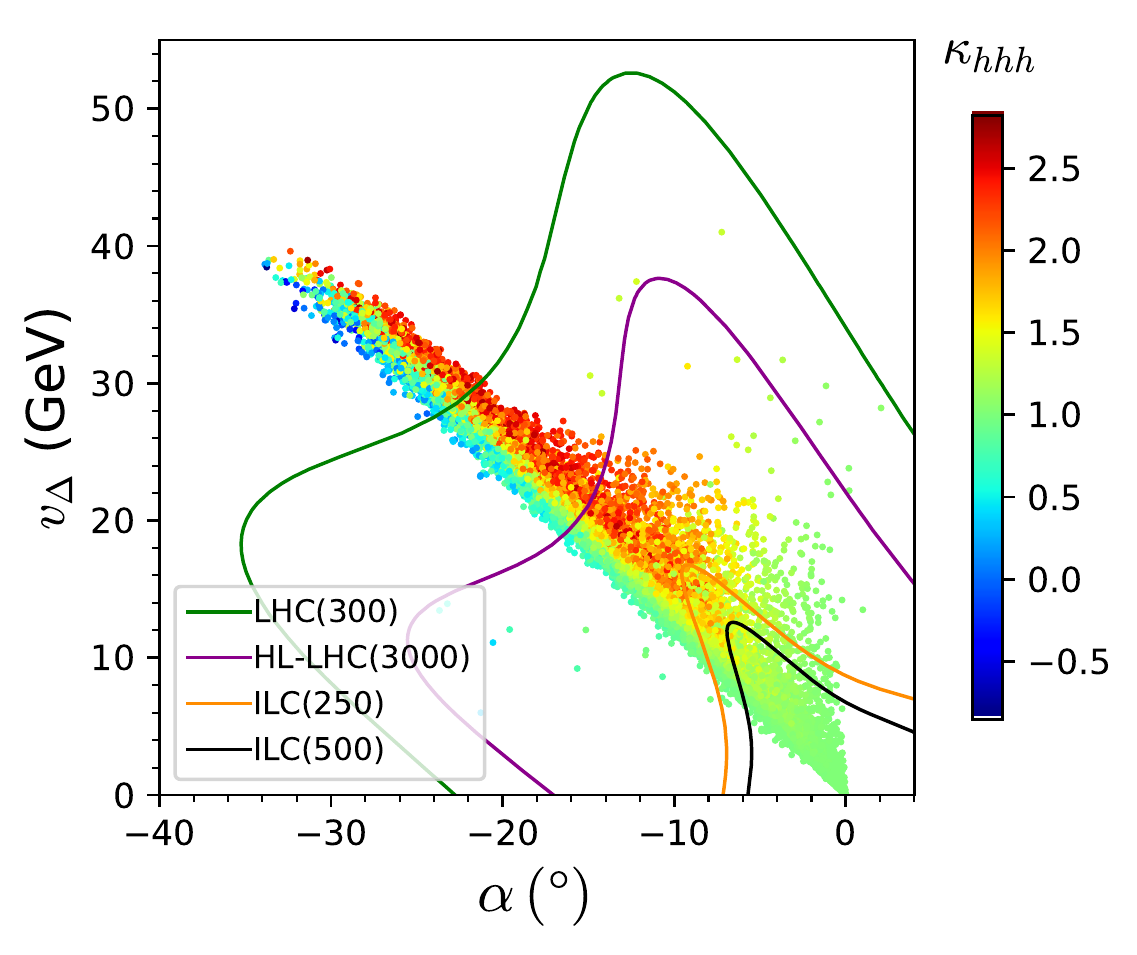}
    \includegraphics[width=0.45\linewidth]{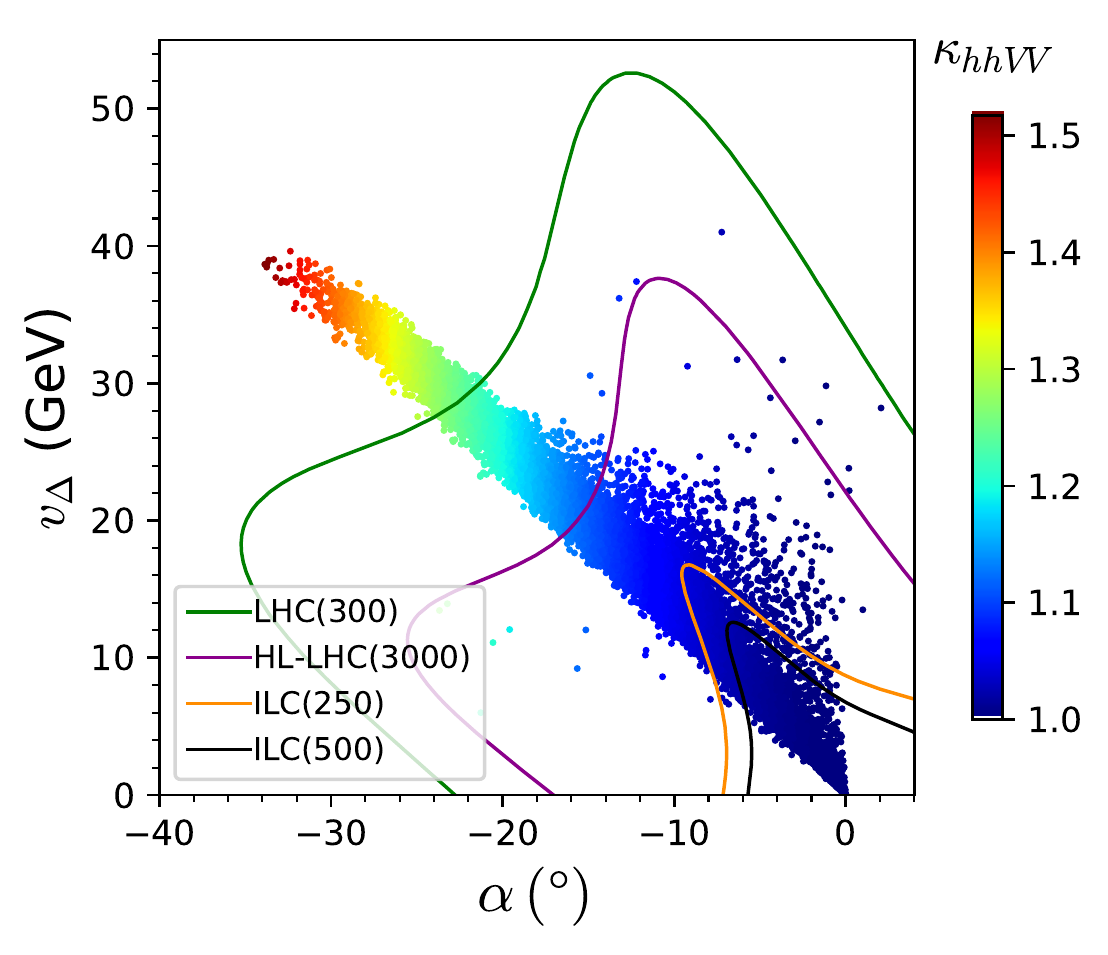}\\
	\includegraphics[width=0.45\linewidth]{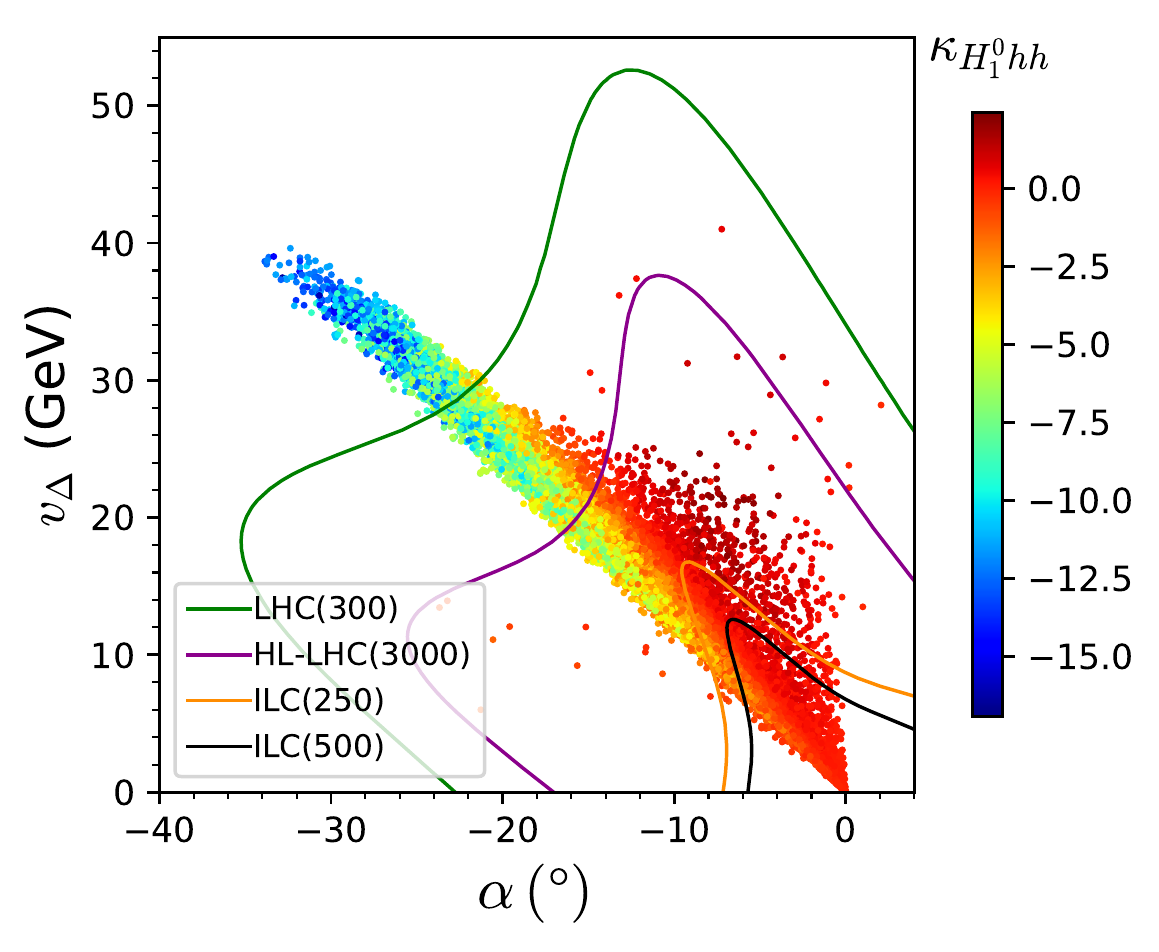}
    \includegraphics[width=0.45\linewidth]{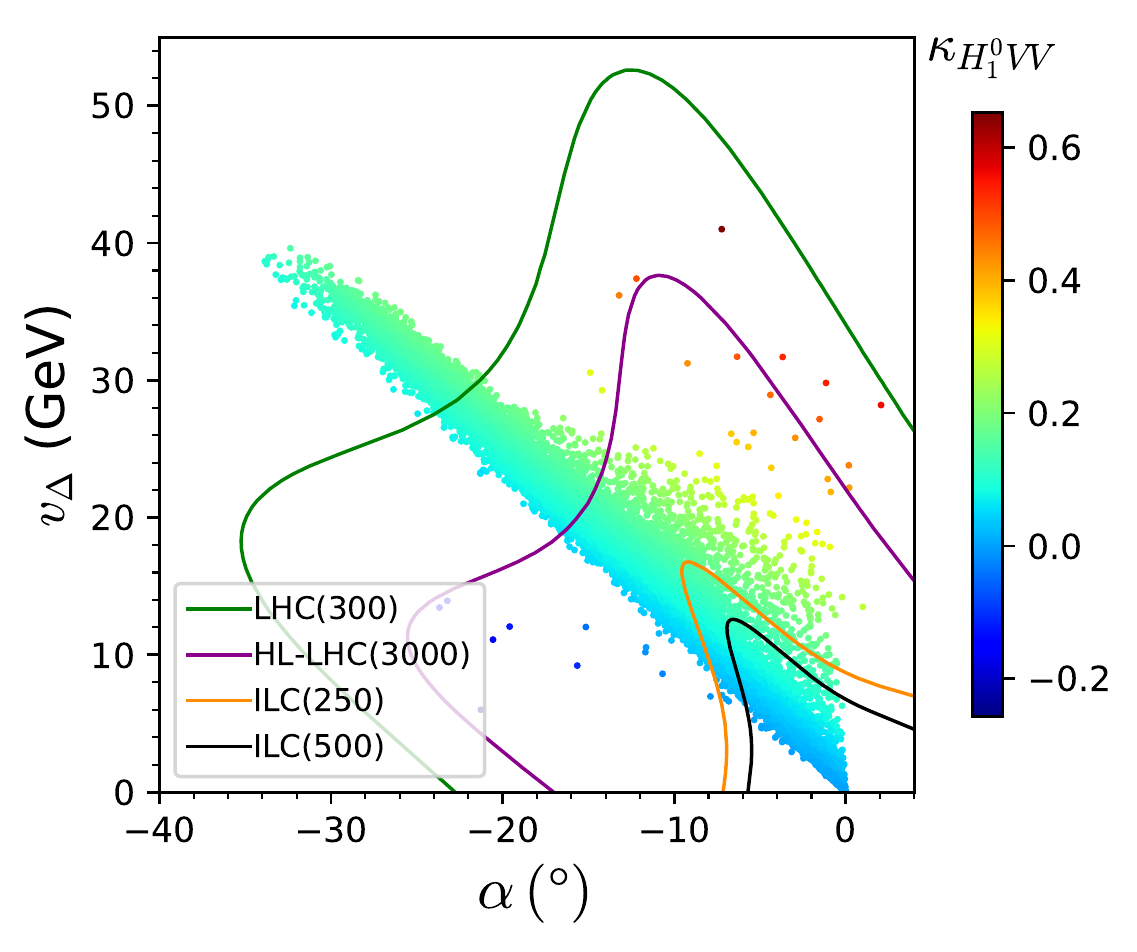}
	\caption[H5]{Same as Fig.~\ref{fig:khtt}, but for $\kappa_{hhh}$, $\kappa_{hhVV}$, $\kappa_{H_1^0hh}$ and $\kappa_{H_1^0VV}$ }
	\label{fig:kH10}
\end{figure}

To study the Higgs self-interactions in the symmetry breaking sector, it is indispensable to measure the Higgs pair production at future $e^+e^-$ colliders. In this subsection we consider possible production mechanisms in the GM model. These processes include the double Higgs-strahlung process $e^+e^-\to hhZ$ in Fig.~\ref{fig:feyn-hhz} and the Higgs pair production via the $W$ boson fusion $e^+e^-\to hh\nu_e\bar{\nu}_e$ in Fig.~\ref{fig:feyn-hhvv} while ignoring the much smaller $Z$ boson fusion~\cite{Abramowicz:2016zbo,Baer:2013cma}. These processes involve the scale factors
\begin{eqnarray}
\kappa_{hhVV}&=&1+\frac{5}{3}s_\alpha^2,
\end{eqnarray}
$\kappa_{H_1^0VV}$ shown in Eq.~(\ref{equ:heavy higgs}), and $\kappa_{hhh}=g_{hhh}/g_{hhh}^{\text{SM}}$, $\kappa_{H_1^0hh}= g_{H_1^0hh}/g_{hhh}^{\text{SM}}$, where $g_{hhh}^{\text{SM}}=3m_h^2/v$, and $g_{hhh},~g_{H_1^0hh}$ are given in Appendix~\ref{app:trilinear}. In Fig.~\ref{fig:kH10}, we show the scanned results for these scale factors in the allowed parameter space. Compared to their SM counterparts, the $hhVV$ coupling is always enhanced due to $\kappa_{hhVV}\geq 1$, whereas the $hhh$ coupling can even change its sign. For those points with $\kappa_{hhh}<0$, LHC@300 will exclude most of them while HL-LHC@3000 could eliminate them altogether. The precision expected to be reachable at the ILC for the measurement of $g_{hhh}$ is of $\mathcal{O}(10\%)$~\cite{Fujii:2015jha}, so the effect of the GM model should be observable. For the new couplings involving $H_1^0$, the magnitude of $\kappa_{H_1^0hh}$ can be so large that the processes are significantly enhanced but perturbation theory may not apply there, while the scale factor $\kappa_{H_1^0VV}$ ranges from $-0.2$ to $0.6$. Considering the limit expected to be available at LHC@300 (HL-LHC@3000), $\kappa_{H_1^0 hh}\lesssim 10$ ($\kappa_{H_1^0 hh}\lesssim 5$) should be satisfied.

\subsubsection{Double Higgs-Strahlung}

The differential cross section for the process $e^+e^-\to hhZ$ can be written as~\cite{Djouadi:1996ah,Osland:1998hv,Djouadi:1999gv}
\begin{align}
\label{equ:sigma-hhz}
\frac{d\sigma(e^+e^-\to hhZ)}{dx_1dx_2}=&
\frac{G_F^3m_Z^6}{384\sqrt{2}\pi^3s}(V_e^2+A_e^2)\frac{\mathcal{A}}{(1-r_Z)^2}.
\end{align}
In the above equation, $x_{1,2}=2E_{1,2}/\sqrt{s}$ with $E_{1,2}$ being the energies of the Higgs bosons, and the shortcuts $x_3=2-x_1-x_2$, $y_i=1-x_i~(i=1,2,3)$, $r_{1,3}=m_{1,3}^2/s$ are used. The function $\mathcal{A}$ can be expressed in the form~\cite{Osland:1998hv}
\begin{align}
\label{eqn:calA}
\mathcal{A}=r_Z\left\{\frac{1}{2}|a|^2f_a+|b(y_1)|^2f_b+2\text{Re}[ab^*(y_1)]g_{ab}
+\text{Re}[b(y_1)b^*(y_2)]g_{bb}\right\}+\{x_1,y_1\leftrightarrow x_2,y_2\},
\end{align}
where
\begin{align}\label{equ:ab}
\nonumber
a=&\frac{1}{2}\bigg(\frac{\kappa_{hVV}\kappa_{hhh}}{y_3+r_Z-\tilde{r}_h}
+\frac{\kappa_{H_1^0VV}\kappa_{H_1^0hh}}{y_3+r_Z-\tilde{r}_1}\bigg)\frac{3m_h^2}{m_Z^2}
+\bigg(\frac{\kappa_{hVV}^2}{y_1+r_h-\tilde{r}_Z}+
\frac{\kappa_{hVV}^2}{y_2+r_h-\tilde{r}_Z}\bigg)+\frac{\kappa_{hhVV}}{2r_Z},\\
b(y)=&\frac{1}{2r_Z}\bigg(\frac{\kappa_{hVV}^2}{y+r_h-\tilde{r}_Z}+
\frac{\kappa_{H_3^0hZ}^2}{y+r_h-\tilde{r}_3}\bigg).
\end{align}
Here $\tilde{r}_i~(i=h,Z,H_1,H_3)$ includes the total decay width of the particle $i$ in its mass squared, i.e., $\tilde{r}_i=(m_i^2-im_i\Gamma_i)/s$ when a resonance is crossed over. The coefficients $f$ and $g$ in Eq.~(\ref{eqn:calA}) are
\begin{align}
\nonumber
f_a=&x_3^2+8r_Z,
\\
\nonumber
f_b=&(x_1^2-4r_h)[(y_1-r_Z)^2-4r_Zr_h],
\\
\nonumber
g_{ab}=&r_Z[2(r_Z-4r_h)+x_1^2+x_2(x_2+x_3)]-y_1(2y_2-x_1x_3),
\\
\nonumber
g_{bb}=&(y_3-x_1x_2-x_3r_Z-4r_hr_Z)(2y_3-x_1x_2-4r_h+4r_Z)
\\
&+r_Z^2(4r_h+6-x_1x_2)+2r_Z(r_Z^2+y_3-4r_h).
\end{align}

\begin{figure}[!htbp]
	\centering
	\includegraphics[width=0.45\linewidth]{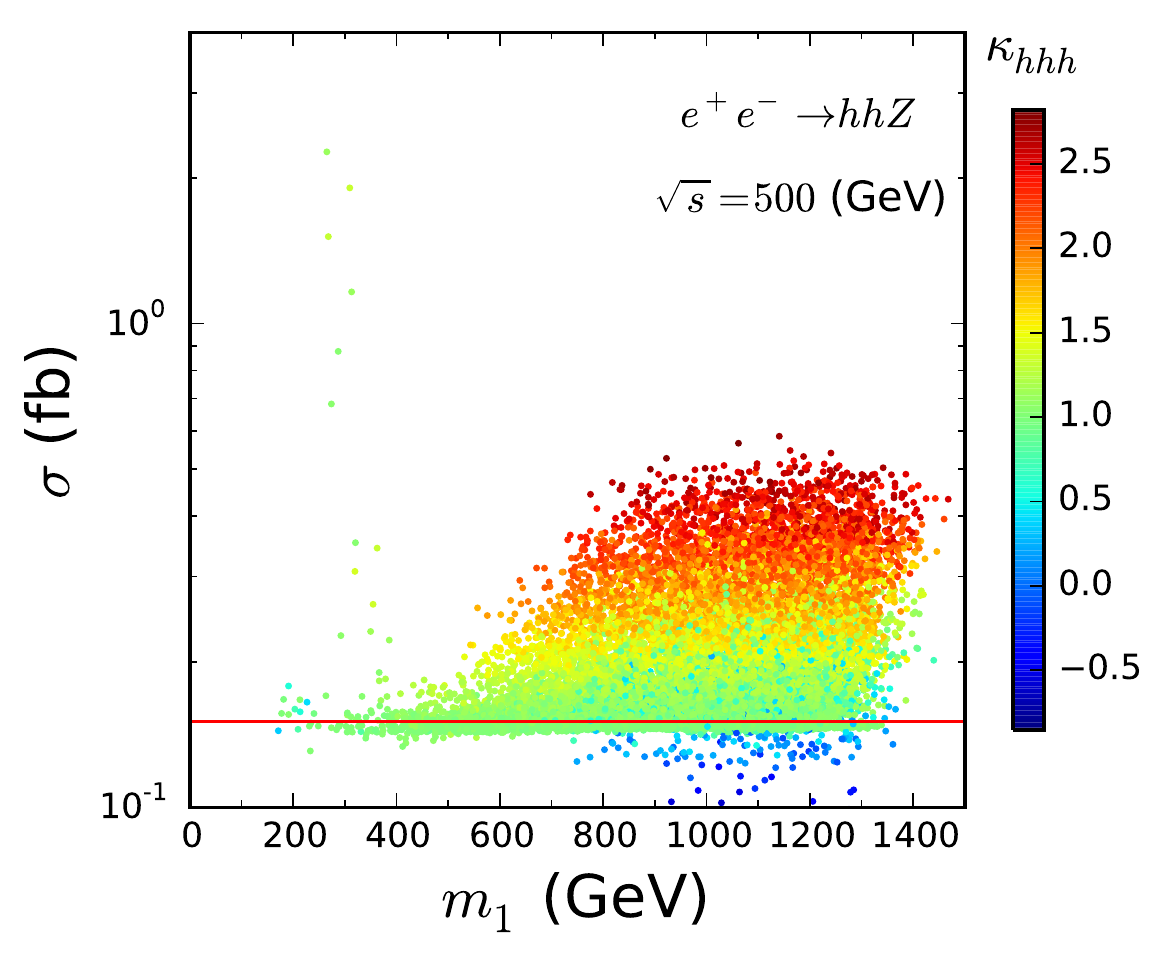}~~
	\includegraphics[width=0.45\linewidth]{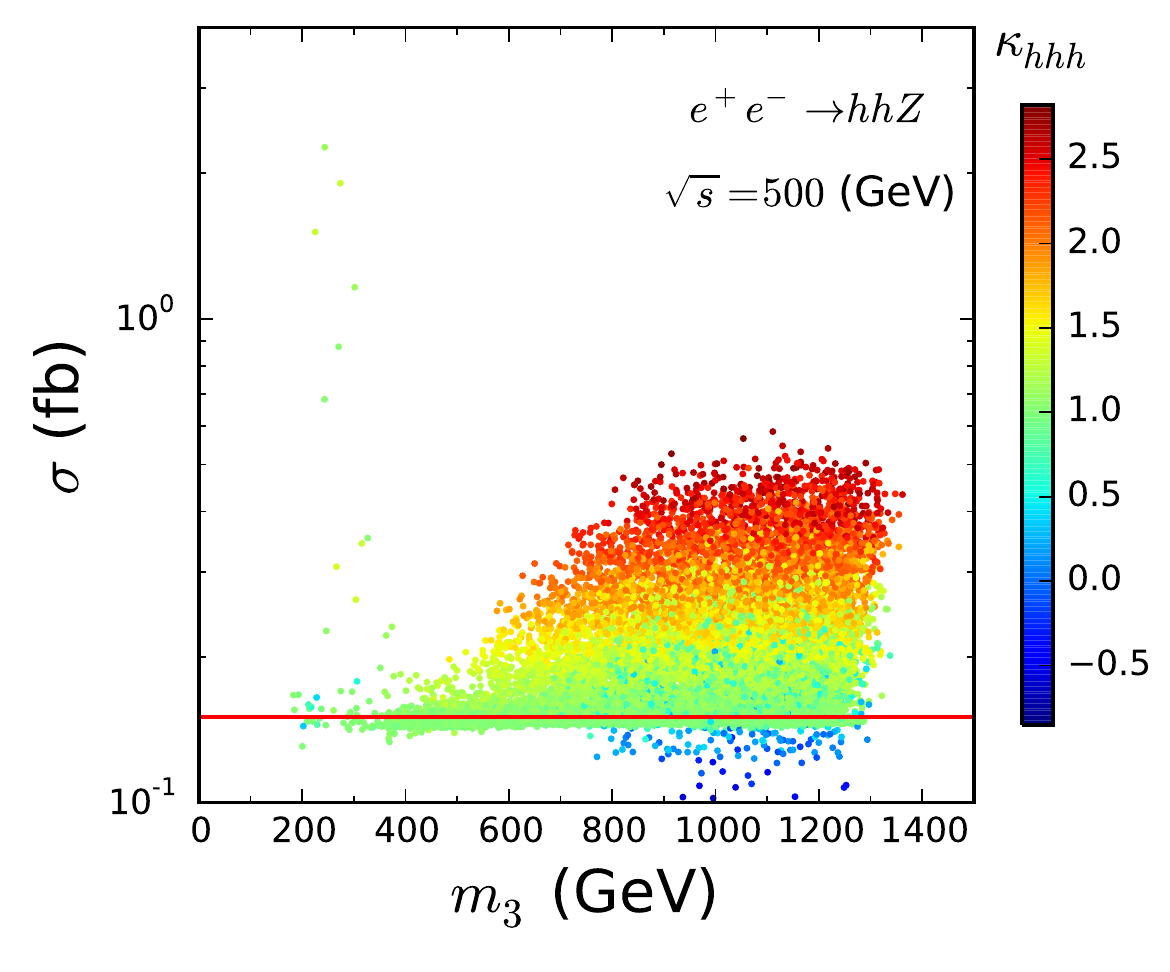}\\
	\includegraphics[width=0.45\linewidth]{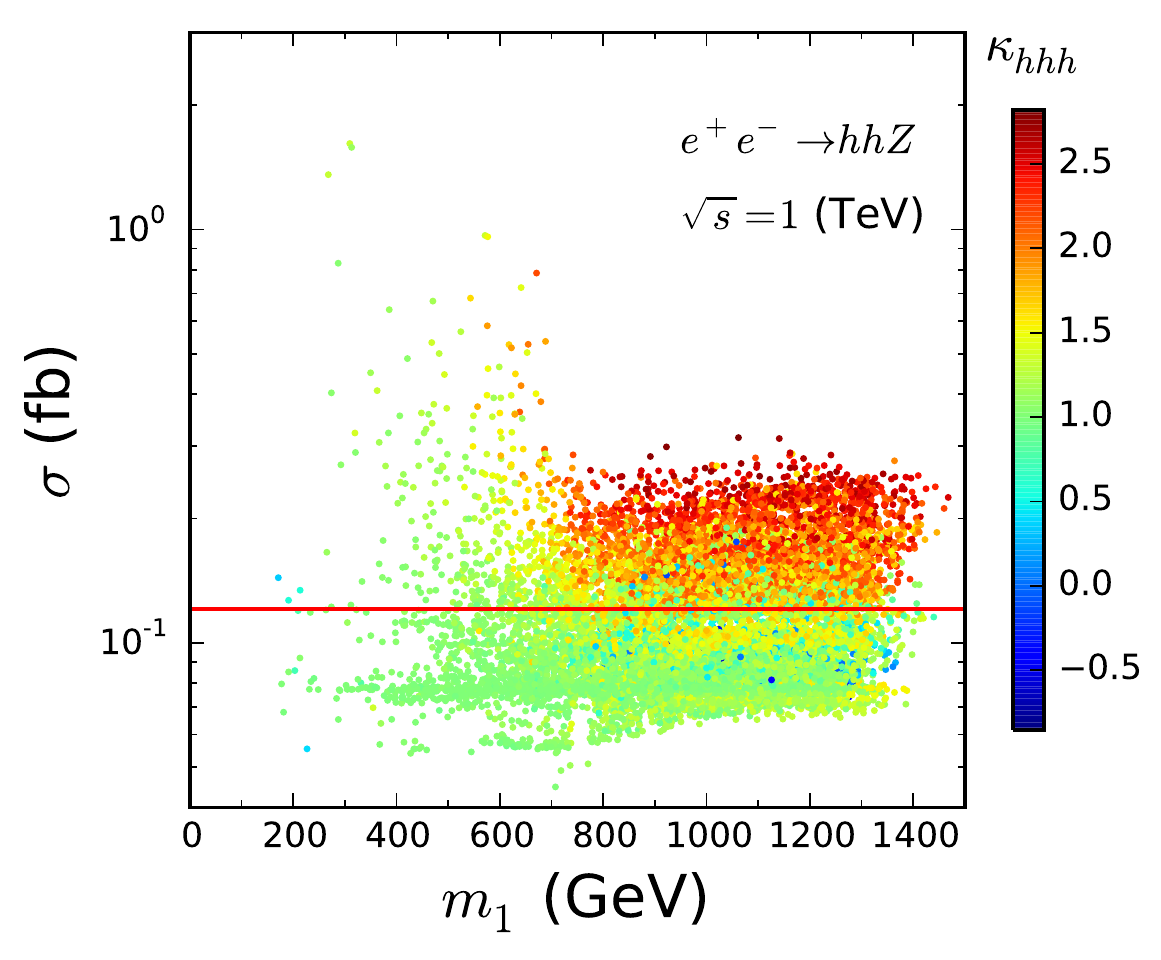}~~
	\includegraphics[width=0.45\linewidth]{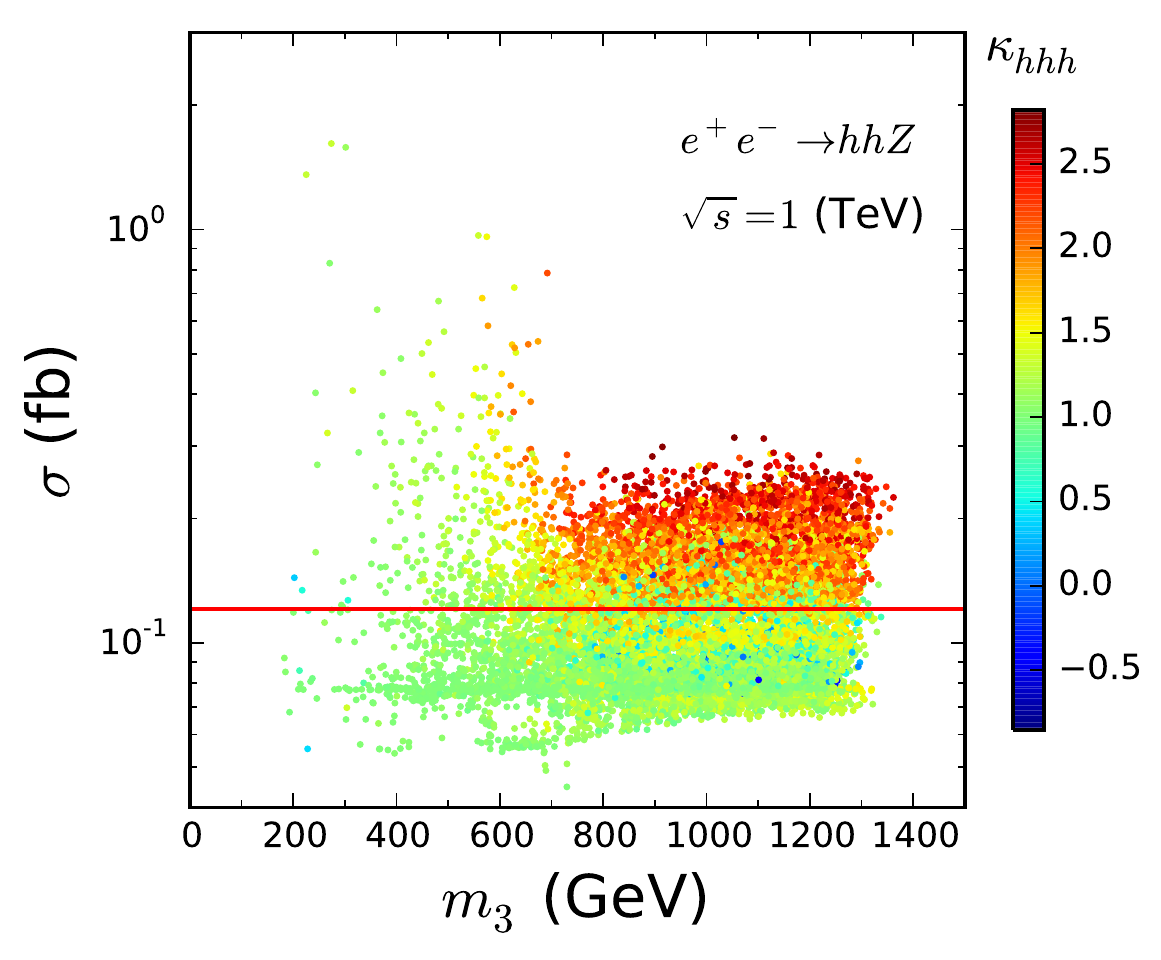}\\
	\caption{Cross section for $e^+e^-\to hhZ$ as a function of the $H_1^0$ mass $m_1$ (left panel) and $H_3^0$ mass $m_3$ (right) at $\sqrt{s}=500~\GeV$ (upper)  and 1~TeV (lower). The horizontal line indicates the SM value.}
	\label{fig:hhz}
\end{figure}

The double Higgs boson production is sensitive to the triple Higgs coupling $g_{hhh}$, which cannot be probed by the single Higgs boson production. In addition, the heavy CP-even Higgs boson $H_1^0$ contributes, thereby enabling sensitivity to the $H_1^0hh$ coupling as well. The total cross section is shown in Fig.~\ref{fig:hhz} at the energy $\sqrt{s}=500~\GeV$ and 1 TeV. Compared to the SM case, the on-shell production of heavy scalars $H_1^0$ and $H_3^0$ followed by decays $H_1^0\to hh$ and $H_3^0\to hZ$ plays a dominant role in the resonant region of the parameter space, where the cross section can increase by more than one order of magnitude.  Since in the resonance region, most of the currently allowed points have $\kappa_{hhh}\approx 1$, the future measurements at LHC@300 and HL-LHC@3000 would be hard to exclude such points. In the non-resonance region, the cross section at $\sqrt{s}=500~\GeV$ is positively correlated with $\kappa_{hhh}$, leading to enhanced cross section with $\kappa_{hhh}>1$ or suppressed cross section with $\kappa_{hhh}<1$. The cross section at $\sqrt{s}=1~\TeV$ can be either enhanced or suppressed even with $\kappa_{hhh}\approx 1$.

\subsubsection{Double Higgs from Vector Boson Fusion}

\begin{figure}[!htbp]
	\centering
	\includegraphics[width=0.9\linewidth]{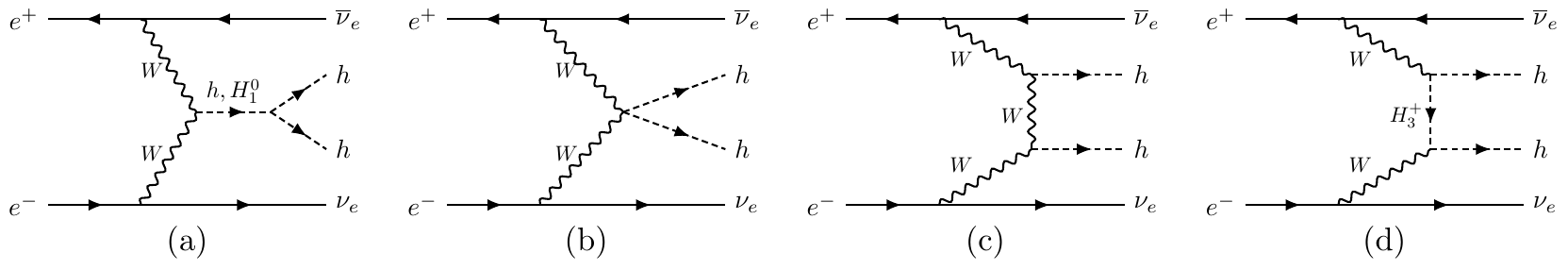}\\
	\caption{Feynman diagrams for double Higgs production from $WW$ fusion at $e^+e^-$ colliders.}
	\label{fig:feyn-hhvv}
\end{figure}

Besides the resonant $WW$ fusion for single $h$ production, there exists non-resonant $WW$ fusion production of a pair of $h$, $e^+e^-\to hh\nu_e \bar{\nu}_e$, shown in Fig.~\ref{fig:feyn-hhvv}. The GM model modifies existing interactions in the SM and introduces new contributions due to $H_1^0$ and $H_3^\pm$ exchanges. The cross section in the effective $W$ approximation can be written as~\cite{Osland:1998hv,Djouadi:1999gv}
\begin{align}
\sigma(e^+e^-\to hh\nu_e\bar{\nu}_e)=\int_{x_\text{min}}^{1}dx\frac{dL}{dx}\hat{\sigma}_{WW}(x),
\end{align}
where $x_\text{min}=4r_h(=4m_h^2/s)$, the differential luminosity function is~\cite{Osland:1998hv}
\begin{align}
\frac{dL(x)}{dx}=\frac{G_F^2m_W^4}{8\pi^4}
\frac{1}{x}[-(1+x)\ln x-2(1-x)],
\end{align}
and the cross section for the subprocess is~\cite{Osland:1998hv}
\begin{align}
\nonumber
\hat{\sigma}_{WW}(x)=&
\frac{G_F^2\hat{s}}{64\pi}\bigg[4\beta_h\Big(
\frac{3r_h\kappa_{hVV}\kappa_{hhh}}{1-r_h}
+\frac{3r_h\kappa_{H_1^0VV}\kappa_{H_1^0hh}}{1-r_1}
+\kappa_{hhVV}\Big)^2
\\
\nonumber
&+2\Big(\frac{3r_h\kappa_{hVV}\kappa_{hhh}}{1-r_h}
+\frac{3r_h\kappa_{H_1^0VV}\kappa_{H_1^0hh}}{1-r_1}
+\kappa_{hhVV}\Big)(\kappa_{hVV}^2F_1+\kappa_{H_3^\pm hW}^2F_2)
\\
&+\beta_h^{-1}(\kappa_{hVV}^4F_3+\kappa_{H_3^\pm hW}^4F_4
+4\kappa_{hVV}^2\kappa_{H_3^\pm hW}^2F_5)\bigg],
\label{eqn:subWW}
\end{align}
where the functions $F_i~(i=1,\dots,5)$ are reproduced in Appendix~\ref{app:hhvv} and $\beta_h=(1-4r_h)^{1/2}$. Note that here $r_i~(i=h,~Z,~H_1,~H_3)$ are defined with respect to $\hat s=xs$, e.g., $r_1=m_1^2/\hat s$.

\begin{figure}[!htbp]
	\centering
	\includegraphics[width=0.45\linewidth]{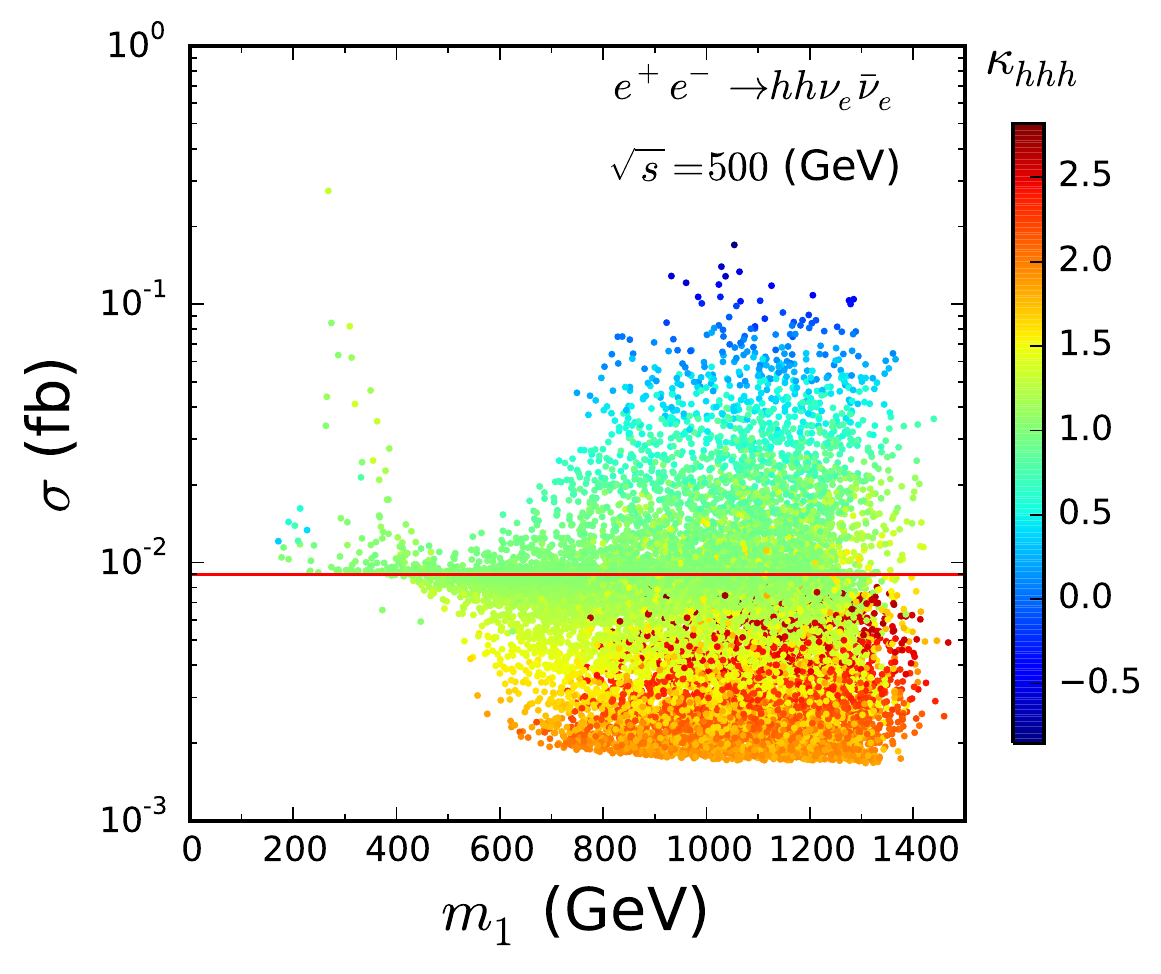}~~
	\includegraphics[width=0.45\linewidth]{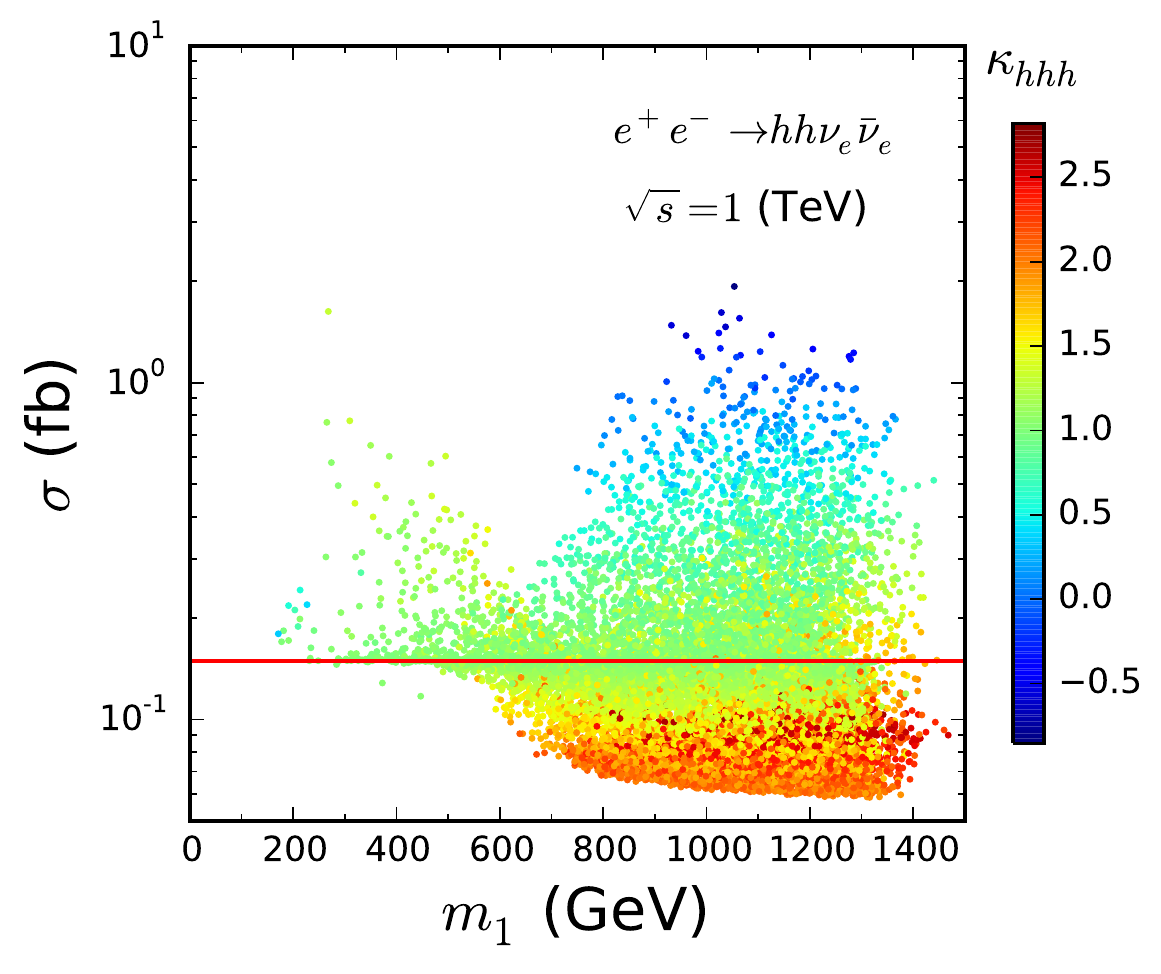}
	\caption{Cross section for $e^+e^-\to hh\nu_e\bar{\nu}_e$ as a function of the $H_1^0$ mass $m_1$ at $\sqrt{s}=500~\GeV$ (left panel) and 1 TeV (right). The horizontal line indicates the SM value.}
	\label{fig:hhvv}
\end{figure}

The total cross section for $e^+e^-\to hh\nu_e\bar{\nu}_e$ is shown in Fig.~\ref{fig:hhvv} at $\sqrt{s}=500~\GeV$ and 1 TeV respectively. It varies in a wide range as $\kappa_{hhh}$ and $\kappa_{H_1^0hh}$ do, indicating its sensitivity to the values of trilinear couplings. Similarly to the case of the double Higgs-strahlung process, the cross section is strongly enhanced in the parameter range where the fusion subprocess, $WW\to H_1^0\to hh$, is resonant also with $\kappa_{hhh}\approx1$. Out of the resonant region the cross section decreases with increasing $\kappa_{hhh}$ due to destructive interference between the scalar and gauge parts of the amplitude.  From Fig.~\ref{fig:kH10}, one sees that $\kappa_{hhh}<1$ would not be favored by HL-HLC@3000. In this case, we expect a smaller cross section in the non-resonance region.

\section{Conclusions}
\label{sec:conclusion}

In this paper, we have investigated the dominant single and double SM-like Higgs ($h$) production at future $e^+e^-$ colliders in the Georgi-Machacek (GM) model. We comprehensively considered various constraints that are currently available, including theoretical, indirect, and direct experimental constraints on the model parameters. In particular, we updated the constraints from searches for heavy singlet ($H_1^0$), triplet ($H_3$), and quintuplet ($H_5$) Higgs scalars and from SM-like Higgs signal strengths with the latest LHC results. The GM model parameter space now becomes more tightly constrained, for instance, only $v_\Delta\lesssim40~\GeV$ with $-30^\circ\lesssim\alpha\lesssim0^\circ$ is allowed now.

Our analysis of the Higgs production has been done in the allowed parameter regions established as above. Besides modifications to the SM couplings, the SM-like Higgs production receives new contributions from new scalars such as $H_1^0,~H_3^0$, and can deviate significantly from the prediction in the SM. Our numerical results have the following key features:

\begin{itemize}
  \item For single Higgs production such as $e^+e^-\to hZ,~h\nu_e\bar{\nu}_e,~he^+e^-$, the production mechanisms are the same as in SM with the only modification occurring in the Higgs-gauge boson coupling $hVV$. Their cross sections are modified by a factor ranging from 0.86 to 1.32.
  \item For associated Higgs production with a top pair $e^+e^-\to ht\bar{t}$, there is a new contribution via $e^+e^-\to hH_3^0$ with $H_3^0\to t\bar{t}$. When $2m_t<m_3<\sqrt{s}-m_h$ is fulfilled, $H_3^0$ becomes resonantly produced, thus enhancing the cross section by up to three times.
  \item For the double Higgs-strahlung process $e^+e^-\to hhZ$, both $H_1^0$ and $H_3^0$ contribute. When resonantly produced, they can lead to an order of magnitude enhancement in the cross section. In the non-resonant region, large deviations are still possible due to a large modification to the trilinear Higgs coupling.
  \item For the double Higgs production via vector boson fusion $e^+e^-\to hh\nu_e\bar{\nu}_e$, both $H_1^0$ and $H_3^\pm$ enter but only $H_1^0$ can be on-shell produced. In the resonant region more than an order of magnitude enhancement is viable, and in the non-resonant region the cross section can maximally increase or decrease by an order of magnitude.
\end{itemize}

Anticipating that LHC will be operated at slightly higher energy and with larger integrated luminosity, we have also estimated its future impact on Higgs couplings and production at electron colliders. We found that LHC@300 and HL-LHC@3000 have the ability to exclude some of the parameter space that is currently allowed. When the new scalars in the GM model are light enough to be resonantly produced at electron colliders, large enhancement in the SM-like Higgs production is possible though its couplings are close to the SM values. In the non-resonance region, the double Higgs production channels are expected to deviate significantly from the SM under the strict future bounds from LHC. If the on-going LHC experiments could observe some deviations of the Higgs couplings, future electron colliders would be capable of confirming it; if not, precise measurements of the Higgs properties at future electron colliders could probe most of the currently allowed parameter space of the GM model. Compared to other popular models such as MSSM and THDM, the GM model can enhance or suppress Higgs couplings to vector bosons and fermions, and drastically change the trilinear Higgs coupling from the SM prediction. The modification of these Higgs couplings leads to distinguished phenomenology for Higgs production processes at $e^+e^-$ colliders. Furthermore, light scalars such as $H_1^0$ and $H_3^0$ can be on-shell produced in certain processes, which provides a good chance to hunt new particles. We thus expect that the model could be tested with high precision and be discriminated from some other new physics scenarios.

\section*{Acknowledgement}

This work was supported in part by the Grants No. NSFC-11575089 and No. NSFC-11025525, by The National Key Research and Development Program of China under Grant No. 2017YFA0402200, and by the CAS Center for Excellence in Particle Physics (CCEPP). We thank the anonymous referee for suggesting the inclusion of projected limits that are expected from future LHC operations.

\appendix

\section{Coefficients for $\sigma(e^+e^-\to ht\bar{t})$}
\label{app:htt}

The seven coefficients $G_i~(i=1,\cdots,7)$ appearing in Eq.~(\ref{equ:htt}) can be obtained by slight adaption from the MSSM case~\cite{Djouadi:1992gp}.
The Higgs radiation from the top quark yields the first two coefficients:
\begin{align}
\nonumber
G_1=&\frac{2g_t^2}{s^2(\beta_t^2-x_h^2)x_h}
\bigg\{-4\beta_t(4m_t^2-m_h^2)(2m_t^2+s)x_h+(\beta_t^2-x_h^2)
\Big[16m_t^4+2m_h^4\\
\nonumber
&-2m_h^2sx_h+s^2x_h^2-4m_t^2(3m_h^2-2s-2sx_h)\Big]
\ln\Big(\frac{x_h+\beta_t}{x_h-\beta_t}\Big)\bigg\},\\
\nonumber
G_2=&\frac{-2g_t^2}{s^2(\beta_t^2-x_h^2)x_h}\bigg\{\beta_t x_h\Big[-96m_t^4+24m_t^2m_h^2-s(m_h^2-s-sx_h)(\beta_t^2-x_h^2)\Big]\\
&+2(\beta_t^2-x_h^2)\Big[24m_t^4+2(m_h^4-m_h^2sx_h)-m_t^2(14m_h^2-12sx_h-sx_h^2)\Big]
\ln\Big(\frac{x_h+\beta_t}{x_h-\beta_t}\Big)\bigg\}.
\end{align}
The other five coefficients all involve the Higgs radiation from the $Z$ boson including its interference with the radiation from the top quark:
\begin{align}
\nonumber
G_3=&\frac{-2\beta_t g_Z^2m_t^2}{m_Z^2(m_h^2-m_Z^2+s-sx_h)^2}
\{4m_h^4+12m_Z^4+2m_Z^2sx_h^2+s^2(x_h-1)x_h^2
\\
\nonumber
&-m_h^2[8m_Z^2+s(x_h^2+4x_h-4)]\},
\\
\nonumber
G_4=&\frac{\beta_t g_Z^2m_Z^2}{6(m_h^2-m_Z^2+s-sx_h)^2}[48m_t^2+12m_h^2+s(24-\beta_t^2-24x_h+3x_h^2)],
\\
\nonumber
G_5=&\frac{4g_{t}g_Zm_t}{m_Zs(m_h^2-m_Z^2+s-sx_h)}\bigg\{\beta_t s\Big[6m_Z^2-x_h(m_h^2+s-sx_h)\Big]
\\
\nonumber
&+2\Big[m_h^2(m_h^2-3m_Z^2+s-sx_h)+m_t^2(12m_Z^2-4m_h^2+sx_h^2)\Big]
\ln\Big(\frac{x_h+\beta_t}{x_h-\beta_t}\Big)\bigg\},
\\
\nonumber
G_6=&\frac{-8g_{t}g_Zm_tm_Z}{s(m_h^2-m_Z^2+s-sx_h)}\bigg[\beta_t s+(4m_t^2-m_h^2+2s-sx_h)\ln\Big(\frac{x_h+\beta_t}{x_h-\beta_t}\Big)\bigg],
\\
\nonumber
G_7=&\frac{-g_{H_3^0t\bar{t}}\kappa_{hH_3^0Z}}{m_h^2-m_{H_3}^2+s-sx_h}
\bigg\{2\beta_t(4m_h^2-sx_h)\Big[\frac{(m_h^2+s-sx_h)g_{H_3^0t\bar{t}}
\kappa_{H_3^0hZ}}{m_h^2-m_{H_3}^2+s-sx_h}-\frac{2m_tg_{Z}}{m_Z}\Big]
\\
&+\frac{2 g_{t}}{s}[2sx_h\beta_t(sx_h-s-m_h^2)-4(m_h^2(sx_h-s-m_h^2)+m_t^2(4m_h^2-sx_h^2))]
\ln\Big(\frac{x_h+\beta_t}{x_h-\beta_t}\Big)\bigg\}.
\end{align}
The relevant $h$ couplings are defined in terms of their $\kappa$ factors:
\begin{align}
g_Z= \frac{m_Z}{v}\kappa_{hVV},~g_{t}= -\frac{m_t}{v}\kappa_{ht\bar{t}},
~g_{H_3^0t\bar{t}}=\frac{m_t}{v}\kappa_{H_3^0 t\bar{t}}.
\end{align}
The kinematical variable $\beta_t$ is
\begin{align}
\beta_t=\bigg\{\frac{[x_h^2-(x_h^\text{min})^2](x_h^\text{max}-x_h)}
{x_h^\text{max}-x_h+4r_t}\bigg\}^{1/2},
\end{align}
where $x_h^\text{min}=2r_h^{1/2}$ and $x_h^\text{max}=1-4r_t+r_h$.

\section{Trilinear Higgs Couplings}
\label{app:trilinear}

The trilinear Higgs couplings used in our calculation include
\begin{align}\label{eq:trilinear}
\nonumber
g_{hhh}=&~24\lambda _1 c_{\alpha }^3 v_{\phi }+\frac{3\sqrt{3}}{2}  c_{\alpha }^2 s_{\alpha } \left[M_1+4\left(\lambda _5-2 \lambda _2\right) v_{\Delta}\right]+6 \left(2\lambda _2-\lambda _5\right) c_{\alpha } s_{\alpha }^2 v_{\phi }\\\nonumber
&+4 \sqrt{3} s_{\alpha }^3\left[M_2-2 \left(\lambda _3+3 \lambda _4\right) v_{\Delta}\right],\\\nonumber
g_{H_1^0hh}=&~4 \left(6\lambda _1-2 \lambda _2+\lambda _5\right) c_{\alpha }^2 s_{\alpha }
v_{\phi } +2 \left(2 \lambda _2-\lambda _5\right) s_{\alpha }^3v_{\phi }\\
&+ \sqrt{3} c_{\alpha } s_{\alpha }^2 \left[M_1-4 M_2+4 \left(2 \lambda _3-2 \lambda_2+6 \lambda _4+\lambda _5\right) v_{\Delta}\right] \\ \nonumber
&-\frac{\sqrt{3}}{2}  c_{\alpha }^3 \left[M_1+4\left(\lambda _5-2 \lambda _2\right) v_{\Delta }\right].
\end{align}
In the decoupling limit, $s_\alpha,v_\Delta,M_1,M_2\to 0$, and $\lambda_1\to m_h^2/(8v^2)$, we have $g_{hhh}\to 24 v\lambda_1\to 3m_h^2/v$ and $g_{H_1^0hh}\to -\sqrt{3}M_1/2\to 0$.

\section{Coefficients for $\sigma(e^+e^-\to hh\nu_e\bar{\nu}_e)$}
\label{app:hhvv}

The coefficients used in the calculation of $e^+e^-\to hh\nu_e\bar{\nu}_e$ are~\cite{Osland:1998hv}
\begin{align}\nonumber
F_1=&~8[2r_W+(r_h-r_W)^2]l_W-4\beta_h(1+2r_h-2r_W),\\\nonumber
F_2=&~8(r_h-r_3)^2l_3-4\beta_h(1+2r_h-2r_w),\\\nonumber
F_3=&~8\beta_h[2r_W+(r_h-r_W)^2][2r_W+1-3(r_h-r_W)^2]\frac{l_W}{a_W}\\\nonumber
&+16[2r_W+(r_h-r_W)^2]^2y_W+16\beta_h^2(1+a_W)^2,\\\nonumber
F_4=&~8\beta_h(r_h-r_3)^2[1-3(r_h-r_3)^2]\frac{l_3}{a_3}\\\nonumber
&+16(r_h-r_3)^4y_3+16\beta_h^2(1+a_3)^2,\\
F_5=&~\frac{\beta_h}{4}(z_Wl_W+z_3l_3)+8\beta_h^2(1+a_W)(1+a_3),
\end{align}
where
\begin{align}
\nonumber
l_{W,3}=&~\ln\frac{1-2r_h+2r_{W,3}-\beta_h}{1-2r_h+2r_{W,3}+\beta_h},\\\nonumber
y_{W,3}=&~\frac{2\beta_h^2}{(1-2r_h+2r_{W,3})^2-\beta_h^2},\\\nonumber
a_{W,3}=&~-\frac{1}{2}+r_h-r_{W,3},\\\nonumber
z_W=&~\frac{(1+2a_W)^2}{a_3-a_W}[8r_W+(1+2a_W)^2]+\frac{(1-2a_W)^2}{a_3+a_W}[8r_W+(1+2a_W)^2],\\
z_3=&~-\frac{(1+2a_3)^2}{a_3-a_W}[8r_W+(1+2a_3)^2]+\frac{(1+2a_3)^2}{a_3+a_W}[8r_W+(1-2a_3)^2].
\end{align}

\bibliographystyle{unsrt}

\end{document}